\documentclass[twocolumn]{raa_twocolumn}

\usepackage{graphicx,grffile,times}
\usepackage{natbib}
\usepackage{amssymb,amsmath}
\bibpunct{(}{)}{;}{a}{}{,}

\usepackage[pagebackref=true]{hyperref}

\usepackage{ulem}

\newcommand{\unit}[1]{{\ \rm #1}}
\newcommand{\txt}[1]{{\rm #1}}

\newcommand{\Mpc}{\ensuremath{\,{\rm Mpc}}}

\newcommand{\Msun}{\ensuremath{\, {\rm M}_{\odot}}}

\newcommand{\HI}{H\textsc{I}}
\newcommand{\HIt}{\textbf{H}\textsc{I}}
\newcommand{\HIs}{\textnormal{H}\textsc{I}}
\newcommand{\HII}{H\textsc{II}}
\newcommand{\SOFIAI}{\textsc{sofia}~1}
\newcommand{\SOFIA}{\textsc{sofia}~2}
\newcommand{\OII}{[O\textsc{II}]}
\newcommand{\OIII}{[O\textsc{III}]}

\graphicspath{{./}{figures/}}

\begin{document}

\title{Cross-Comparison of Galaxies Detected in the CSST Spectroscopic Survey and the SKA \HIt\ Survey}

\volnopage{ {\bf 20XX} Vol.\ {\bf X} No. {\bf XX}, 000--000}
\setcounter{page}{1}

\author{Yingfeng Liu \inst{1,2}, Furen Deng\inst{1,2}, Wenxiang Pei\inst{7}, Haitao Miao\inst{1}, Qi Xiong\inst{1,2}, Shuanghao Shu\inst{1,2}, Xingchen Zhou\inst{1,2}, Qi Guo\inst{3,4,5}, Yan Gong\inst{5,2}, Yougang Wang\inst{1,2}, Xuelei Chen\inst{1,2,6,*} \footnotetext{$*$Corresponding Author}}

\institute{State Key Laboratory of Radio Astronomy and Technology, National Astronomical Observatories, Chinese Academy of Sciences, Beijing 100101, China, {\it xuelei@cosmology.bao.ac.cn}\\
\and
School of Astronomy and Space Science, University of Chinese Academy of Sciences, Beijing 100049, China\\
\and
Institute for Frontiers in Astronomy and Astrophysics, Beijing Normal University, Beijing 102206, People’s Republic of China\\
\and
School of Physics and Astronomy, Beijing Normal University, Beijing 100875, People’s Republic of China\\
\and
Key Laboratory for Computational Astrophysics, National Astronomical Observatories, Chinese Academy of Sciences, Beijing 100101, China\\
\and
Center of High Energy Physics, Peking University, Beijing 100871, China\\
\and
Shanghai Key Lab for Astrophysics, Shanghai Normal University, Shanghai 200234, People’s Republic of China\\
\vs \no
{\small Received 20XX Month Day; accepted 20XX Month Day}
}

\abstract{
We present a forward-modeling framework to forecast the galaxies detected in the Chinese Space Station Survey Telescope (CSST) spectroscopic survey and the Square Kilometre Array (SKA) \HI\ survey.
Starting from the L-Galaxies 2020 semi-analytic model run on the Millennium-II N-body simulation (MS-II), the cold gas in galaxies is partitioned into atomic and molecular components self-consistently within the model. We further model the emission-lines (H\,$\alpha$, H\,$\beta$, \OIII) relevant for the slitless spectrograph of the CSST in a post-processing step.
We construct mock lightcones using the Mock Map Facility (MoMaF) approach, simulating the neutral hydrogen (\HI) data cubes representing a $2000$ hour SKA-Mid spectral line observation from redshifts $0.25$--$0.5$, and employ the Source Finding Application 2 (\SOFIA) source-finding package to generate an \HI\ galaxy catalog.
In parallel, we apply the CSST selection function and noise model to obtain a realistic catalog of emission-line galaxies; the emission-line signal is proportional to the star formation rate.
These products allow us to cross compare the galaxy samples and assess the synergy between CSST and SKA.
We study the correlations of the \HI\ and the emission-line signal with the halo mass, \HI\ mass, and the stellar mass, and the baryonic Tully-Fisher relation (BTFR).
We also perform stacking analysis of the \HI\ signal from the CSST-selected sample, which probes the \HI\ content in galaxies with low \HI\ mass.
Finally, we derive the optical-\HI\ cross-correlation power spectrum of the galaxies, and measure the bias of these galaxies.
These results can provide useful insight on the cold gas and stellar content of the galaxies.
}

\authorrunning{Y.-F. Liu et al. }    
\titlerunning{CSST-SKA HI galaxies}  
\maketitle

\section{Introduction}\label{sec:intro}

Cross comparison of observations with different probes is a powerful technique that has become increasingly important in modern astrophysics\citep{Alonso2015,Pourtsidou2016,Cohn2016,Chensf2019,Witzemann2019,Shi2020,Padmanabhan2020,Modi2021}.
Observations in distinct wavelength regimes often trace different physical processes or galaxy populations; therefore, combining them offers a more comprehensive understanding of cosmic structures and their evolution, revealing physical connections between these populations and providing complementary constraints on both astrophysical and cosmological parameters.
Cross comparison of optical spectroscopic galaxy surveys with \HI\ galaxy or intensity mapping surveys provides a sensitive probe of large-scale structure and galaxy evolution.
This approach has been widely explored in recent years~\citep{Wolz2016, Cunnington2023, Jiang2023, Deng2022}.
The cross-correlation of different bands mitigates selection biases and survey-specific systematics, such as instrumental noise and residual foreground contamination, while offering joint constraints on cosmology and galaxy formation~\citep{Cunnington2019, Wolz2017}.

The Chinese Space Station Survey Telescope (CSST) is designed to conduct a wide-area galaxy survey, including a multi-color photometric survey and a survey with slitless spectroscopy~\citep{Zhan2011SSPMA, Zhan2018,Zhan2021,Gong2019,Gong2026,Gong2025}.
The latter will detect large numbers of emission-line galaxies (ELGs) via their H\,$\alpha$, H\,$\beta$, \OIII, and \OII\ lines in the redshift range $0.2 \lesssim z \lesssim 1.6$, and measure their redshifts spectroscopically with high accuracy.
On the other hand, the Square Kilometer Array (SKA) is the largest radio telescope project.
The SKA-Mid array will map the cosmic distribution of neutral hydrogen \HI\ via the 21\,cm line, either through detection of a large number of \HI\ galaxies, or through \HI\ intensity mapping of the large scale structure~\citep{groupCosmologyPhaseSquare2020}.
Both the CSST and the SKA-Mid are expected to come online in the later 2020s.
The combination of CSST and SKA thus provides a unique opportunity to study the link between galaxies and their \HI\ reservoirs.

To establish a realistic connection between theoretical predictions and forthcoming observations, it is crucial to construct mock lightcones from cosmological simulations.
Unlike static snapshots that represent the Universe at a single cosmic epoch, lightcones reproduce the continuous evolution of structure along the observer’s line of sight, thereby mimicking the observational perspective of surveys such as CSST and SKA-Mid.
This approach enables a consistent treatment of redshift-dependent effects, selection functions, and survey geometry, all of which are essential for meaningful cross-correlation studies.
Moreover, lightcone-based mock catalogs and data cubes provide a natural framework for incorporating observational systematics—such as flux limits, instrumental beams, and noise—allowing direct comparison with real survey data.
By capturing cosmic evolution in a geometrically realistic way, lightcone simulations serve as indispensable tools for testing analysis pipelines, interpreting observational signals, and forecasting the scientific return of future multi-wavelength surveys.

In this work we present the analysis of the cross-correlation between optical galaxies observed by the CSST and \HI\ galaxies observed by the SKA-Mid.
This is based on the following simulation pipeline:
\begin{enumerate}
    \item Run the L-Galaxies 2020 semi-analytic model~\citep{Henriques2020} on a high-resolution cosmological N-body simulation to obtain galaxy populations with physical properties.
    \item Compute emission-line fluxes in a post-processing step, while the \HI\ content is obtained self-consistently from the semi-analytic model.
    \item Construct lightcones using the Mock Map Facility (\texttt{MoMaF}) method~\citep{Blaizot2005}.
    \item Simulate SKA \HI\ data cubes including instrumental effects, and apply \SOFIA\,\footnote{\url{https://gitlab.com/SoFiA-Admin/SoFiA-2}}~\citep{Westmeier2021} to obtain an observed catalog.
    \item Apply CSST instrumental effects and selection criteria to generate an ELG catalog.
\end{enumerate}

We describe the details of our methodology in the following sections.
In Section~\ref{sec:simu}~we introduce the simulations that underpin our work, and describe how the \HI\ content is obtained from the semi-analytic model and the emission-line properties are assigned in post-processing.
Section~\ref{sec:mock_obs}~explains the construction of the mock lightcone using the MoMaF approach, including the treatment of box replication and redshift-space distortions, and how the simulated lightcones are converted into mock observations: the generation of SKA \HI\ data cubes, the source extraction with \SOFIA, and the construction of the CSST emission-line galaxy catalog with realistic instrumental effects and selection functions.
In Section~\ref{sec:result}~we present the results for the galaxy samples detected by the SKA \HI\ galaxy survey and the CSST emission-line galaxy survey, including the distributions of halo mass, stellar mass, \HI\ mass, and star formation rate (SFR).
We also make a statistical analysis of the cross-correlation signal between the CSST galaxies and the SKA \HI\ galaxies.
Finally, we conclude in Section~\ref{sec:conclusion}.

\section{Simulations}\label{sec:simu}

This section describes the numerical simulations and theoretical frameworks employed in generating our optical galaxy catalog and \HI\ intensity maps.
Specifically, we utilize the semi-analytic galaxy catalog from~\citet{Henriques2020}.
The galaxy catalog is generated by running the galaxy formation model \textsc{L-Galaxies} onto merger trees extracted from dark matter only $N$-body cosmological simulation Millennium-II~\citep{Boylan-Kolchin2009}.

\subsection{N-body simulation}\label{sec:nbody}

We use the Millennium-II simulation as the basis of our work.
This is a high-resolution cosmological N-body simulation that assumed the WMAP first-year cosmology~\citep{Springel2005b}, with the spacing given by
\begin{equation}\label{eq:wmap_z}
    \log_{10}(1 + z_{N}) = \frac{N(N+35)}{4200} \qquad (0 \leq N \leq 64)
\end{equation}
In this work, we use the MS-II that is rescaled to \emph{Planck} first-year cosmology parameters \citep{Henriques2015,Planck2013} using the method of~\citet{Angulo2010}, which adjusts both the simulation length scale and the snapshot redshifts to match the linear growth of structure in the target cosmology.
The cosmological parameter values are summarized in Table~\ref{tab:spec_params}.
This simulation tracks $2160^3 \sim 10^{10}$ particles from $z \sim 56.4$ to $0$, and the particle mass is $7.69 \times 10^6 M_{\odot}/h$ and the size of the periodic box is $96.06\,h^{-1}\,{\rm Mpc}$.
Following~\cite{Springel2005b}, dark matter halos and subhaloes are identified using the friends-of-friends (FOF) and SUBFIND~\citep{White2001} algorithms.
The smallest halo traced has 20 particles.
This N-body simulation result is used as the basis for the semi-analytical model.
\begin{table}
\centering
 \caption{Cosmological parameters}\label{tab:spec_params}
 \begin{tabular}{cc}
\hline\hline
Parameter & Value\\
\hline
        $L_{\text{box}}$ & $96.0558 \, \Mpc/h$ \\
        $N_p$ & $2160^3$ \\
        $M_p$ & $7.69 \times 10^6 M_{\odot}/h$ \\
        $h$ & 0.673 \\
        $\Omega_M$ & 0.315 \\
        $\Omega_B$ & 0.0488 \\
        $\Omega_{\Lambda}$ & 0.685 \\
        $n_s$ & 0.96 \\
        $\sigma_8$ & 0.829 \\
\hline        
\end{tabular}
\end{table}

\subsection{Semi-analytic model}\label{sec:himodel}

We implement the L-Galaxies semi-analytic model following the framework presented in~\citep{Henriques2020}.
This model incorporates comprehensive physical prescriptions for key baryonic processes, including but not limited to shock heating, radiative gas cooling, star formation regulation through supernova feedback, supermassive black hole (SMBH) formation and growth, active galactic nuclei (AGN) feedback mechanisms, and chemical enrichment processes.
Specifically,~\citet{Henriques2020}~divides the gas and stellar discs into a series of concentric rings and considers all related physical processes in each ring.
They also consider the inflow and outflow between rings.
In this way, they could model the spatial distribution of the stellar and gas discs of galaxies.

In this framework, the amount of cold gas is determined self-consistently by following the exchange of mass between reservoirs: hot halo gas cools radiatively and is deposited into a cold disc with an exponential surface-density profile set by the halo angular momentum, while cold gas is depleted through star formation and reheated or ejected by stellar/AGN feedback.
Cold gas is tracked on a ring-by-ring basis, with radial inflow redistributing mass between rings.

The molecular hydrogen fraction in each ring is not a free parameter but is computed self-consistently from the cold gas surface density and metallicity using a shielding model~\citep{Krumholz2009, McKee2010, Fu2013}.
The hydrogen surface density is given by $\Sigma_{\rm H}=0.74\,\Sigma_{\rm gas}$, with the factor $0.74$ accounting for the hydrogen mass fraction, the remainder being helium and metals~\citep{Obreschkow2009}.
Following their prescription, the molecular and atomic components of the cold gas are given by
\begin{equation}
\Sigma_{\rm H_2} = f_{\rm H_2}\,\Sigma_{\rm H}, \qquad
\end{equation}
\begin{equation}
\Sigma_{\rm HI} = (1-f_{\rm H_2})\,\Sigma_{\rm H}.
\end{equation}
with the molecular-gas fraction $f_{\mathrm{H}_2}$ given by
\begin{align}
f_{\mathrm{H}_2} = 
\left\{
  \begin{array}{ll}
    1 - \frac{0.75 s}{1+0.25s}, & s < 2 \\
    0, & s \geq 2
  \end{array}
\right.
\end{align}
The dimensionless shielding parameter $s$ is given by
\begin{equation}
s = \frac{\ln(1+0.6\chi+0.01\chi^2)}{0.6\,\tau_c}
\end{equation}
in which
\[
  \chi = 0.76 \left(1+3.1\,Z'^{0.365}\right), \qquad \tau_c = 0.066 \, \Sigma_{\mathrm{comp}}\,Z',
\]
where $Z'$ is the metallicity relative to the solar value, and $\Sigma_{\mathrm{comp}}$ is the gas surface density of the gas cloud.
A clumping factor $c_f$ is adopted to model the non-uniform distribution of gas in real galaxies.
\begin{equation}
\Sigma_{\rm comp} = c_f \, \Sigma_{\rm gas}
\end{equation}
This formulation ensures that dense, metal-rich regions become H$_2$-dominated, while diffuse, metal-poor regions remain \HI-dominated.
\begin{figure}[t!]
\centering
\includegraphics[width=0.98\linewidth]{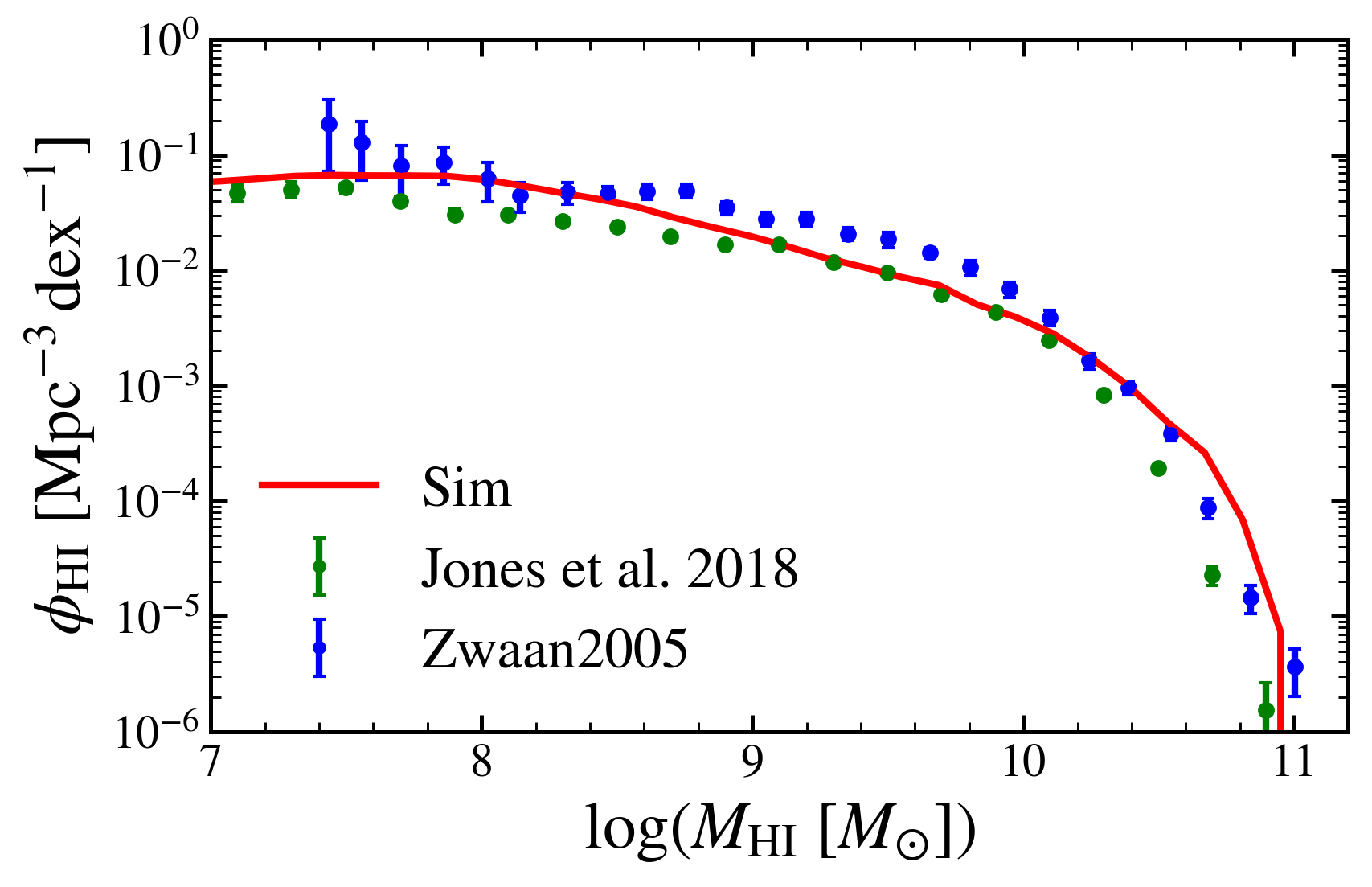}
\caption{Comparison of simulation results for the the \HI\ mass function at $z = 0$ with observational data from~\cite{Zwaan2005} and~\cite{Jones2018}.\label{fig:HIMF_snap}}
\end{figure}

In Figure~\ref{fig:HIMF_snap}, we compare the simulated H{\sc i} mass function at $z=0$ (solid red line) with observational measurements from HIPASS~\citep{Zwaan2005} and ALFALFA~\citep{Jones2018}.
The simulation is in good agreement with the data across the full observed mass range.
\begin{figure}[th]
\centering
\includegraphics[width=0.98\linewidth]{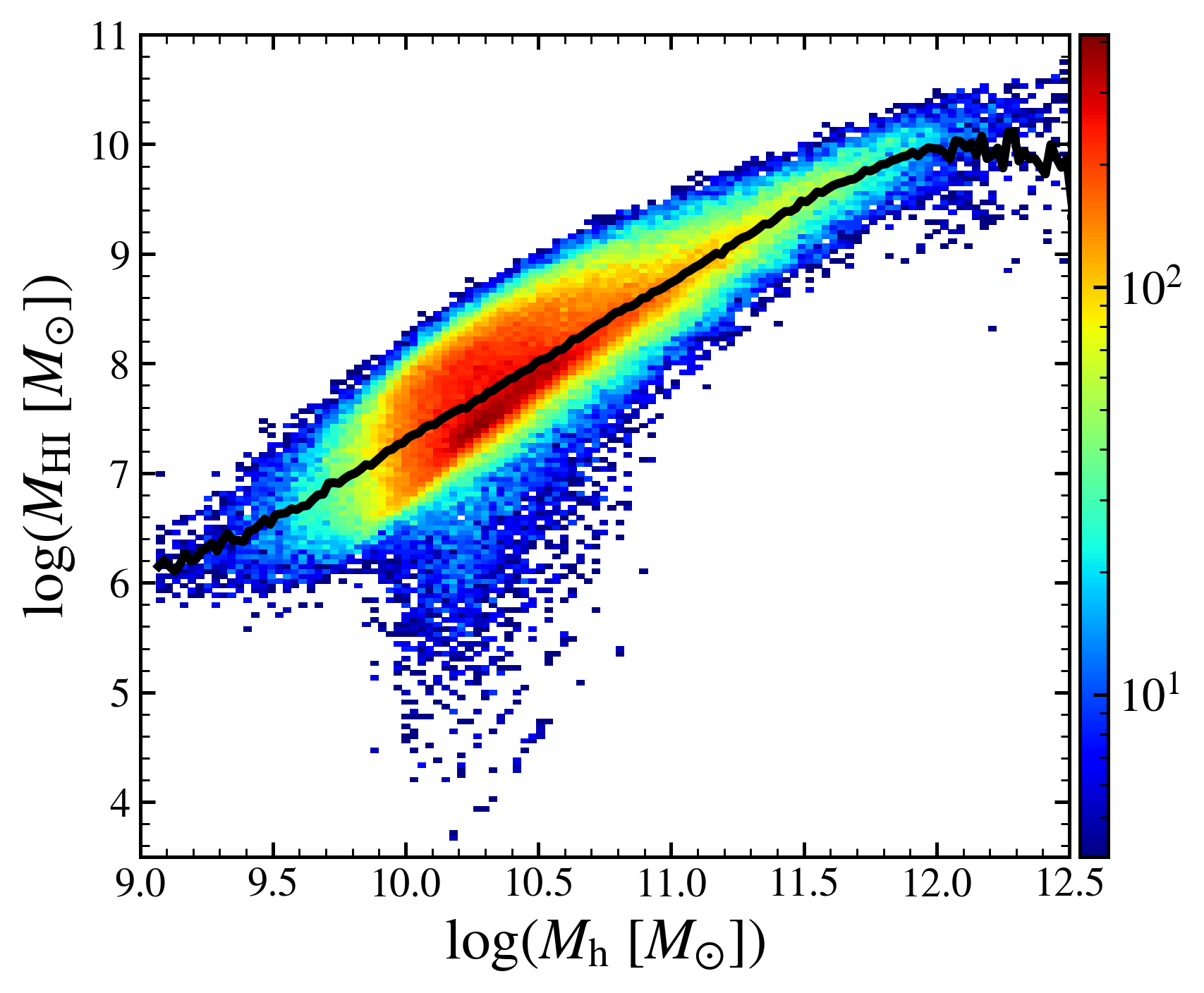}
\caption{The total \HI\ mass as a function of the halo mass at $z=0$, the colorbar shows the number counts in logarithmic scale. The black solid line represents the median $\log(M_{\rm HI})$ in bins of halo mass, computed using fixed-width bins in $\log(M_{\rm h})$.}\label{fig:MHi_Mh_density}
\end{figure}

Figure~\ref{fig:MHi_Mh_density} shows the distribution of total neutral hydrogen mass $M_{\rm HI}$ as a function of halo mass $M_{\rm h}$ at $z=0$ in the simulation, color-coded by number of halos on a logarithmic scale in the $\log M_{\rm h}$–$\log M_{\rm HI}$ plane.
The \HI\ mass of each halo is obtained by summing over all galaxies (centrals, satellites, and orphans) within the halo, while the halo mass given is the one associated with the central galaxy.
Only halos with $M_{\rm h} > 100 \times m_{\rm p}$ (where $m_{\rm p}$ is the simulation particle mass) are included to ensure numerical reliability.
Overall, $M_{\rm HI}$ increases with halo mass, with substantial scatter at fixed $M_{\rm h}$; at the high-mass end, the relation flattens, indicating slower growth of \HI\ content in massive halos due to feedback.

The star formation surface density is assumed to be proportional to the $\rm H_2$ surface density,
\begin{equation}
\Sigma_{\rm SFR}(r) = \alpha_{\rm H_2}\,\frac{\Sigma_{\rm H_2}(r)}{t_{\rm dyn}(r)},
\end{equation}
where $\alpha_{\rm H_2}$ is a calibrated efficiency parameter, and $t_{\rm dyn}(r)$ is a local dynamical timescale defined from the cold disc radius and the halo maximum circular velocity.
The star formation rate (SFR) is thus tied explicitly to the availability of molecular hydrogen, and the resulting SFR is an emergent quantity rather than a free parameter~\citep{Fu2013,Henriques2015,Henriques2020}.

\subsection{The galaxy emission-line model}\label{sec:emimodel}

The CSST slitless spectrometer has low spectral resolution and is mainly sensitive to galaxy emission lines.
Based on the semi-analytic model above, we further compute the luminosities of various emission lines by post-processing following~\citet{Pei2024a}, which adopts the same BC03 SPS model~\citep{Bruzual2003} in both galaxy SED and nebular emission.

The H\,$\alpha$ luminosity can be expressed in terms of the ionizing photon production rate as
\begin{equation}\label{eq:halpha}
L(\mathrm{H}\alpha) = \epsilon_{\mathrm{H}\alpha}\, Q_{\mathrm{H}},
\end{equation}
where
\begin{equation}
\epsilon_{\mathrm{H}\alpha} = \frac{\alpha_{\mathrm{H}\alpha}^{\mathrm{eff}}}{\alpha_{\mathrm{B}}}\, h \nu_{\mathrm{H}\alpha}
\simeq 1.37 \times 10^{-12}\ \mathrm{erg}
\end{equation}
is the energy emitted in H\,$\alpha$ per ionizing photon under case B recombination.
Here, $\alpha_{\mathrm{H}\alpha}^{\rm eff}$ is the effective recombination coefficient for H\,$\alpha$, and $\alpha_{\rm B}$ is the case B recombination coefficient.
The ionizing photon rate $Q_{\mathrm{H}}$ is related to the SFR by
\begin{equation}\label{eq:sfr_q}
Q_{\rm H} = 1.35 \times 10^{53}\ \left(\frac{\rm SFR}{M_\odot / \mathrm{yr}}\right)\ \mathrm{s^{-1}},
\end{equation}
assuming the Chabrier initial mass function (IMF, \citealt{Chabrier2003}).

We also adopt the emission lines generated by \citet{Pei2024a}.
\citet{Pei2024a} utilized the radiative transfer code \texttt{CLOUDY}~\citep{Ferland2017} to calculate the relative strengths of various emission lines relative to H\,$\alpha$ for \HII\ regions with different parameters.
They adopt local empirical relations to calculate the ionization parameter and hydrogen density of \HII\ regions from the general properties of galaxies.
For a given galaxy, the ratio of each emission line to H\,$\alpha$ is derived by interpolating within a pre-computed grid of line ratios, and the luminosities of various emission lines are then obtained.

We note that in this framework the emission lines are assumed to originate solely from stellar photoionization in \HII\ regions, and the contribution from AGN is not explicitly included.
In more comprehensive models, emission lines may also arise from AGN narrow-line regions, which can enhance high-ionization lines (e.g. \OIII) and modify line ratios (e.g. \citealt{EuclidCollaboration2024, Hirschmann2023}).
The omission of AGN emission may therefore lead to an underestimation of the luminosities of some emission lines, particularly at the bright end, and could affect the predicted detection rate of emission-line galaxies.
However, since the CSST emission-line sample is expected to be dominated by star-forming galaxies, the overall impact on the total detection rate is likely to be moderate.

\section{Mock Observations}\label{sec:mock_obs}

In this section, we give a detailed description of the method implemented to obtain a galaxy catalog that can be observed by the CSST optical survey, and a \HI\ galaxy catalog that can be detected from a 2000-hour SKA-Mid mock observation.

\subsection{Lightcone construction}\label{sec:lightcone}

To enable direct comparison with real survey data, it is essential to construct lightcones rather than rely on static simulation snapshots.
Unlike a single-epoch snapshot, a lightcone captures the continuous cosmic evolution of structures and galaxies along the line of sight, incorporating redshift-dependent effects and survey geometry.
We now outline the method to construct the lightcones, which follows the \texttt{MoMaF} method developed by~\citet{Blaizot2005} and~\citet{Kitzbichler2007}.

The simulation box side-length, $L_{\rm box} = 96.06\,h^{-1}\,{\rm Mpc}$, corresponds to a comoving distance out to $z \simeq 0.03$ from $z=0$ in the adopted cosmology.
To generate a cosmological volume of larger size, we stack replications of the simulation box along and transverse to the line-of-sight.
The number of replications per axis, $n_{\rm rep}$, that need to be stacked around the original box is given by
\begin{equation}
n_{\rm rep} = \lfloor \frac{r_{\max}}{L_{\rm box}} \rceil + 1,
\end{equation}
where $r_{\max}$ is the maximum comoving distance to be reached in the final mock catalog.
However, it also introduces a number of artifacts.
Because the same structures reappear periodically, galaxies or clusters may be artificially repeated in the mock cone.
When projected on to the sky, these repeated structures can appear at small angular separations, biasing clustering measurements.~\citet{Blaizot2005} showed that such replication effects tend to suppress the correlation signal on scales below $\sim 0.1 L_{\rm box}$, leading to a negative bias in the recovered clustering amplitude.

A common strategy to mitigate these artifacts is to apply random tiling, in which replicated boxes are transformed by random rotations, translations or reflections before being placed in the lightcone~\citep{Blaizot2005, Henriques2012}.
This helps decorrelate the appearance of repeated structures.
Nevertheless, random tiling has the drawback that, because of the periodic boundary 
conditions of the underlying $N$-body simulation, any transformation other than 
pure translation introduces discontinuities in the density field.
As a result, the underlying large-scale clustering can be distorted, 
which is problematic for statistical analyses that require a faithful representation of the underlying density field.

In addition to replication artifacts, mock lightcones also suffer from 
finite-volume effects~\citep{Merson2013}.
Since the simulation box is finite, 
it cannot capture density fluctuations on scales larger than $L_{\rm box}$.
This results in the absence of large-scale modes in the power spectrum, which can bias clustering 
statistics and introduce sample variance in the mock catalogs.
Therefore, while tiling is essential to reach large survey volumes, 
the combined effects of random tiling and finite box size must be carefully 
accounted for when interpreting clustering measurements from lightcone mocks.
\begin{figure}[th]
\centering
\includegraphics[width=0.98\linewidth]{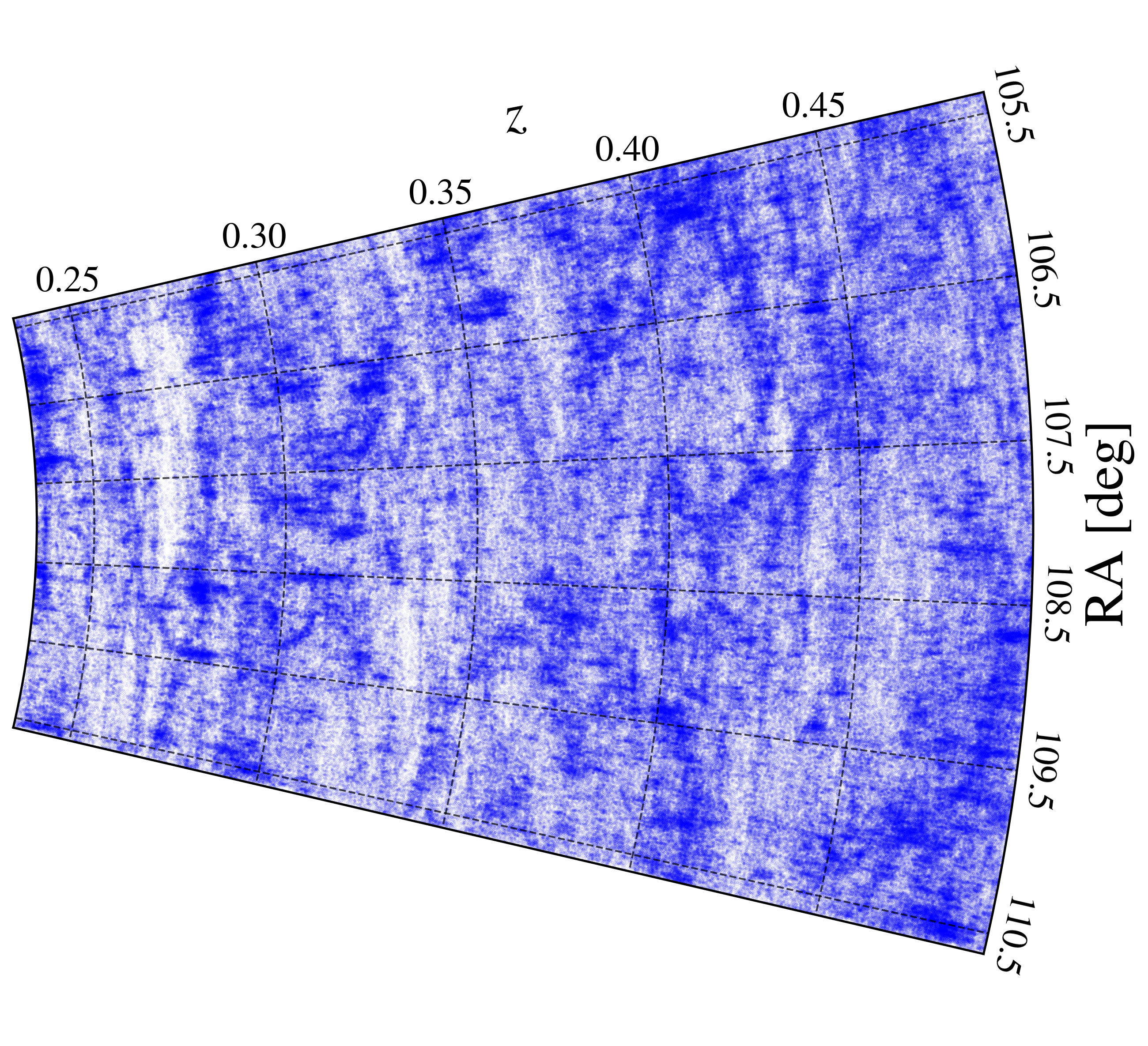}
\caption{The distribution of all galaxies in the lightcone.}\label{fig:lightcone}
\end{figure}

In the \texttt{MoMaF} procedure, the past lightcone of an observer is filled by galaxies drawn from the simulation snapshots.
Taking into account the resolution limit of the simulation, we select galaxies with stellar mass $M_{\ast} > 10^7\,M_{\odot}/h$ in each snapshot.
For each redshift interval, we select the snapshot closest in cosmic time, so that the intrinsic galaxy properties are assigned consistently with their evolutionary stage~\citep{Henriques2012, Merson2013}.
To cover the desired solid angle, the simulation cube is periodically replicated.
In order to mitigate obvious repetition of large-scale structures, random 
rotations and translations are applied to the replicated cubes, following 
\citet{Blaizot2005}.
The observer is placed at the origin, and the Cartesian 
coordinates of galaxies in each replicated cube are shifted according to the 
replication numbers $(n_x,n_y,n_z)$, giving super-box coordinates 
\((X,Y,Z)=(x + n_x L_{\rm box}, \,y + n_y L_{\rm box}, \, z + n_z L_{\rm box})\).

These positions are then converted into observed sky coordinates 
(right ascension and declination) and comoving distance.
The comoving distance to a galaxy is
\[
  r = \sqrt{X^2+Y^2+Z^2},
\]
which is mapped to a cosmological redshift, $z_{\cos}$, via
\begin{equation}
r = \int_0^{z_{\cos}} \frac{c}{H(z')}\,dz' .
\end{equation}
and redshift-space distortions (RSD) are incorporated by adding the 
line-of-sight component of the peculiar velocity $v_{\parallel}$ to the 
cosmological redshift.
The observed redshift is then
\begin{equation}
z_{\rm obs} = z_{\cos} + (1+z_{\cos})\frac{v_\parallel}{c}
\end{equation}
where $v_{\parallel}$ is measured along the direction of the observer.
This procedure ensures that the final lightcone not only includes the  evolution of galaxy properties with redshift but also reproduces the  anisotropies in the observed distribution due to peculiar motions.
The resulting lightcone constitutes a continuous mock observation extending to a redshift limit of $z = 0.5$.
Figure~\ref{fig:lightcone} presents the distribution of galaxies within the lightcone over the redshift range $0.235 < z < 0.495$.

\subsection{The SKA-Mid data cube}\label{sec:IM}

We construct a realistic \HI\ imaging data cube from the lightcone catalogue, incorporating observational noise characteristics following the procedure developed for the SKA Science Data Challenge~2\,\footnote{\url{https://sdc2.skao.int/}} (SDC2). We utilize the SKAO-SDC2 pipeline\,\footnote{\url{https://github.com/PhilippaHartley/SKAO-SDC2}} with bug fixes and MPI parallelization for improved computational efficiency.
Below we summarize the procedure for generating the \HI\ and noise cubes; further details can be found in~\citet{Hartley2023}.

\subsubsection{\HIs\ sky model}

The \HI\ mass $M_{\rm HI}$ of each source in the lightcone catalogue is converted to an integrated line flux $F$ using the relation from~\citet{Duffy2012}:
\begin{equation}
    \frac{F}{\mathrm{Jy\,Hz}} = \frac{1}{49.8}
    \frac{M_{\rm HI}}{M_{\odot}} 
    {\left( \frac{D_{L}(z)}{\Mpc} \right)}^{-2},
\end{equation}
where $D_{L}(z)$ is the luminosity distance.
We impose a lower flux limit of $F_{\rm min} = 1~\mathrm{Jy\,Hz}$, ensuring that a face-on, unresolved source at this threshold produces a peak flux density comparable to the noise r.m.s.\ level.

The semi-analytic models do not generate a spatial resolved \HI\ distribution on galaxy scale, therefore we take advantage of the high-quality observed \HI\ sources to mock the spatial and morphological distribution for model galaxies.
Each galaxy in the catalog is matched to an observed \HI\ template from an atlas of 55 well-resolved galaxies compiled from the THINGS~\citep{Walter2008} and HALOGAS~\citep{Heald2011} surveys.
These templates provide three-dimensional \HI\ data cubes with realistic morphologies and kinematics.
For each simulated galaxy, a subset of candidate templates is identified based on proximity in a parameter space defined by \HI\ mass and inclination angle, and one template is randomly selected from this subset. 
This preserves the correspondence between the simulated and observed galaxy properties while avoiding repeated use of identical morphologies.

The selected template is then positioned and rescaled to match the target right ascension, declination, and frequency.
Scaling factors for the \HI\ major axis ($D_{\rm HI}$), minor axis ($b$), and line width ($w_{20}$) are determined from the galaxy's \HI\ mass, inclination angle $i$, and redshift $z$. Before resampling, a low-pass filter was applied to suppress high-frequency structures and prevent aliasing artifacts.
The target source is then rotated to match the catalog position angle and normalized to the required integrated flux $F$. Each galaxy $i$ thus makes a contribution of specific spectral intensity  $I_i(\alpha,\delta,\nu)$ to the image cube.
The intrinsic (noise-free) \HI\ sky cube is constructed by summing all galaxy contributions:
\begin{equation}
    I_{\rm sky}(\alpha,\delta,\nu) = \sum_{i} I_i(\alpha,\delta,\nu).
\end{equation}
This naturally accounts for source overlap and line-of-sight blending.

\subsubsection{Instrumental response and noise}

\begin{table}
\centering
\caption{The Adopted SKA-Mid Survey parameters}\label{tab:SKA_survey_param}
\begin{tabular}{ll}
\hline\hline
{\bf Parameter} & {\bf Value}\\
Total Integration Time & 2000 hour\\
Field of View & 20 ${\deg}^2$\\
Frequency Coverage & $950$--$1150$ MHz \\
Redshift & $0.235$--$0.495$\\
Beam Width & 7 arcsec \\
Sensitivity & $26$--$31$ $\mu$Jy/beam\\
\hline
\end{tabular}
\end{table}

For the adopted observational setup, the synthesized beam of the SKA-mid is well-approximated by a circular Gaussian with FWHM $= 7''$, and the thermal noise is generated self-consistently through the interferometric simulation in SDC-2 \citep{Hartley2023}. Each frequency channel produces a noise realization with r.m.s.\ level determined by the uv-sampling and flagged visibilities, rescaled for the expected integration time.
For a total observation time of 2000~h, this yields an r.m.s.\ sensitivity of $26$–$31~\mu\mathrm{Jy\,beam^{-1}}$ per channel.
The final observed data cube combines the beam-convolved sky signal with the noise cube:
\begin{equation}
    I_{\rm obs}(\alpha,\delta,\nu) = I_{\rm beam}(\alpha,\delta,\nu) + N(\alpha,\delta,\nu),
\end{equation}
where $N(\alpha,\delta,\nu)$ is assembled from the independent noise realizations for each channel.

The resulting cube spans the frequency range $950$–$1150~\mathrm{MHz}$ (corresponding to $z \approx 0.235$–0.495) with a channel width of $30~\mathrm{kHz}$, and covers a field of view of $\sim 20~\deg^2$ with a synthesized beam of $7''$ sampled on $2.8''$ pixels. 
This corresponds to rest-frame velocity resolutions of $7.8$–$9.5~\mathrm{km\,s^{-1}}$ across the redshift range. 
The adopted survey parameters are summarized in Table~\ref{tab:SKA_survey_param}.

\subsection{\HIt\ source detection}

\begin{figure*}[ht!]
  \centering
  \includegraphics[width=0.98\linewidth]{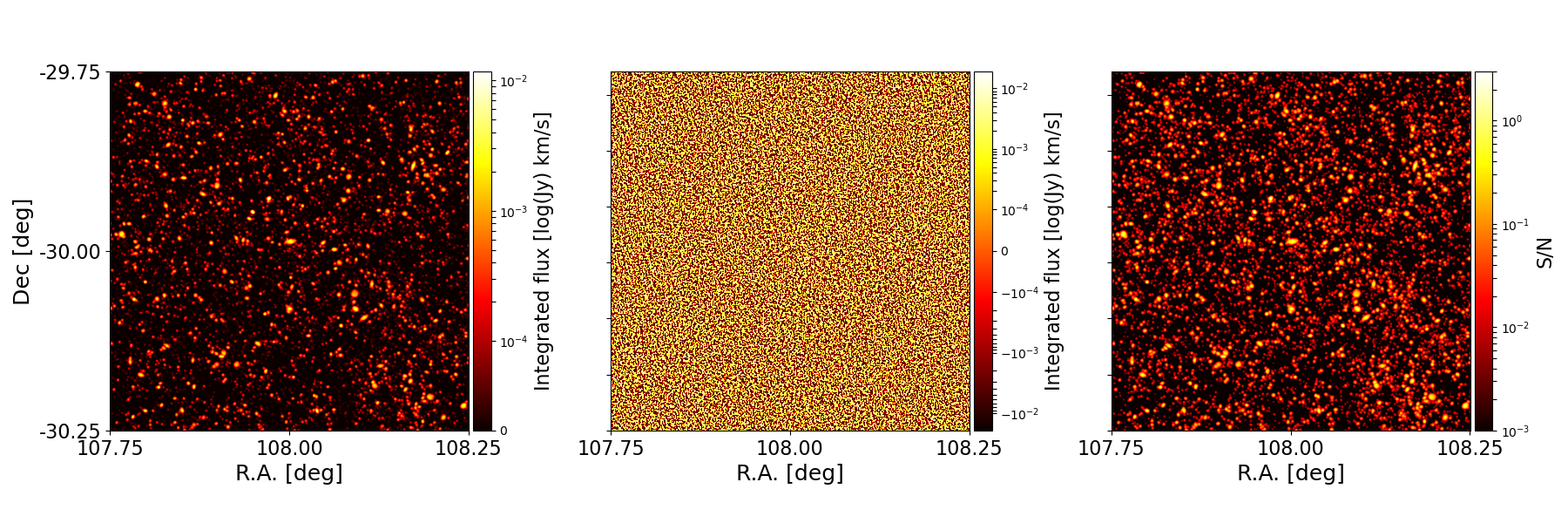}
  \caption{Left panel: \HI\ cube; Middle panel: noise cube; Right panel: SNR.}\label{fig:HI-noise-SNR}
\end{figure*}

The resulting noisy \HI\ datacube was processed with the \SOFIA\ source finder to identify and characterize \HI\ detections, from which the SKA-Mid galaxy catalog was constructed.
Figure~\ref{fig:HI-noise-SNR} shows the images of the \HI\ cube, noise cube and the signal-to-noise ratio (SNR) for a small ($0.25$ square degrees) field of view (FoV).
We performed blind source finding on the three-dimensional SKA-Mid data cubes using the \SOFIA\ (\textit{Source Finding Application 2}) package\,\footnote{\url{https://gitlab.com/SoFiA-Admin/SoFiA-2}}.
\SOFIA\ is a reimplementation of \SOFIAI, incorporating key algorithms such as the automatic radio-frequency interference (RFI) flagging, noise normalization, the S+C source finder, and reliability estimation.
The package supports parallel processing, enabling efficient source extraction from large spectral datasets, which is well suited for the \HI\ line data in this study.
\begin{figure*}[htbp]
    \centering
    \includegraphics[width=0.9\textwidth]{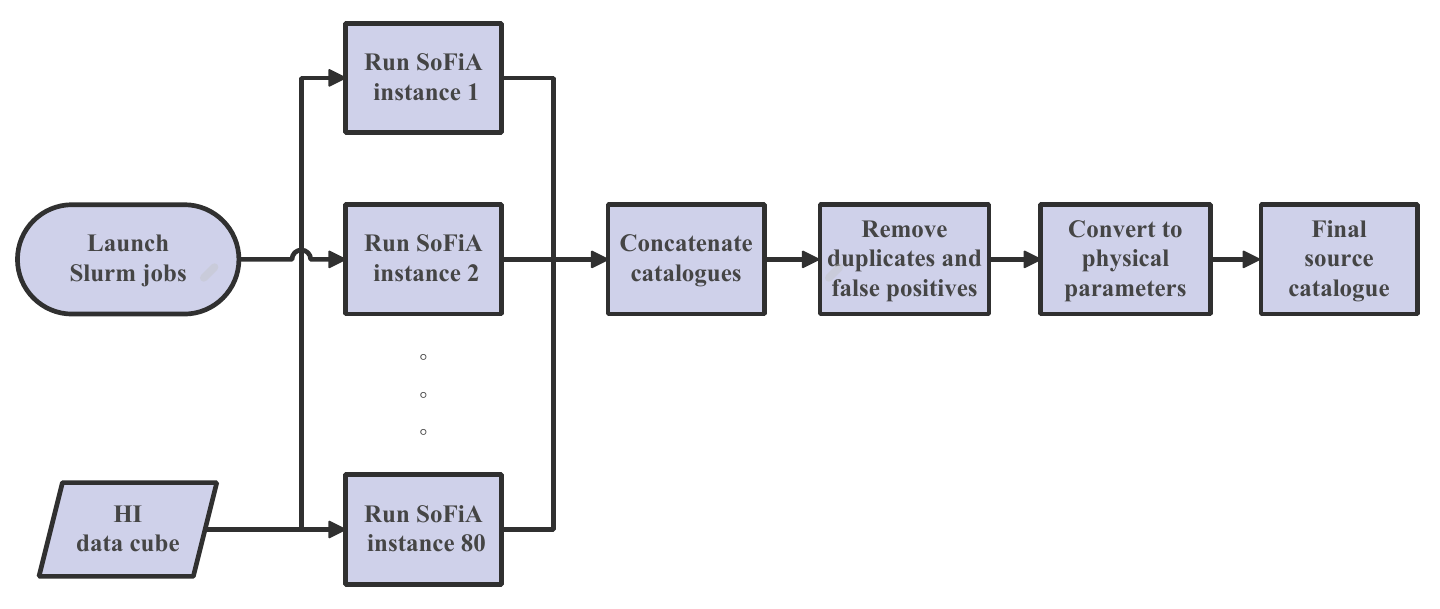}
    \caption{Workflow of blind source finding using \SOFIA.\label{fig:source_find_flow_chart}}
\end{figure*}

We optimized the \SOFIA\ input parameters by exploring the parameter space via a grid search, with the aim of maximizing the source detection performance.
The procedure is as follows:
\begin{enumerate}
    \item Parameter lists were defined for each key module, including \texttt{replacement} and \texttt{minSizeZ} in \texttt{scfind}, \texttt{radiusXY} and \texttt{radiusZ} in \texttt{linker}, and \texttt{minSNR}, \texttt{threshold}, and \texttt{scaleKernel} in \texttt{reliability}.
    \SOFIA\ was run in parallel on a development dataset covering $0.25$\,deg$^2$ to obtain detection results for each parameter combination.
    \item Source catalogs from each run were compared with the true sources in the development set.
Performance metrics included the total number of detections, match rate (ratio of true detections to total detections), and a composite score.
The best-performing parameter set was selected.
    \item To ensure robustness, the evaluation was repeated across multiple subcubes.
The final parameter set was chosen based on consistent performance across all subregions and is given in Table~\ref{tab:sofia_params}.
\end{enumerate}
The standard \SOFIA\ workflow is illustrated in Fig.~\ref{fig:source_find_flow_chart}.
The optimal parameters listed in Table~\ref{tab:sofia_params} were used to run \SOFIA: noise normalization was applied to each spectral channel; the S+C finder used a $3.5\,\sigma$ detection threshold, with spatial kernels of 0, 3, and 7 pixels and spectral kernels of 0, 3, 7, 15, 21, and 31 channels.
The linking radius was 2 pixels/channel with a minimum size of 3 pixels/channel.
Reliability filtering was applied with a threshold of 0.1, a minimum SNR of 1.5, and a kernel scale factor of 0.3.
\begin{table}
\centering
\caption{Optimal parameter settings for \SOFIA}\label{tab:sofia_params}
\begin{tabular}{ll}
\hline\hline
\textbf{Module} & \textbf{Parameters and Values} \\
\hline
\texttt{scaleNoise} & \texttt{windowXY} = 25 \\
                    & \texttt{windowZ} = 15 \\
\texttt{scfind} & \texttt{replacement} = 1.0 \\
                & \texttt{threshold} = 3.5 \\
                & \texttt{kernelsXY} = [0, 3, 7] \\
                & \texttt{kernelsZ} = [0, 3, 7, 15, 21, 31] \\
\texttt{linker} & \texttt{radiusXY/Z} = 2 \\
                & \texttt{minSizeXY/Z} = 3 \\
\texttt{reliability} & \texttt{threshold} = 0.1 \\
                     & \texttt{scaleKernel} = 0.3 \\
                     & \texttt{minSNR} = 1.5 \\
                     \hline
                     
\end{tabular}
\end{table}

The full data cube was divided into 80 overlapping subcubes (with a 64-pixel margin) to enable parallel processing and to avoid truncating sources at the boundaries. 
The resulting source catalogues from each subcube were merged into a single catalogue, with derived quantities converted to physical units for consistency with the input mock catalogue.
\begin{figure}[ht]
\centering
\includegraphics[width=0.95\linewidth]{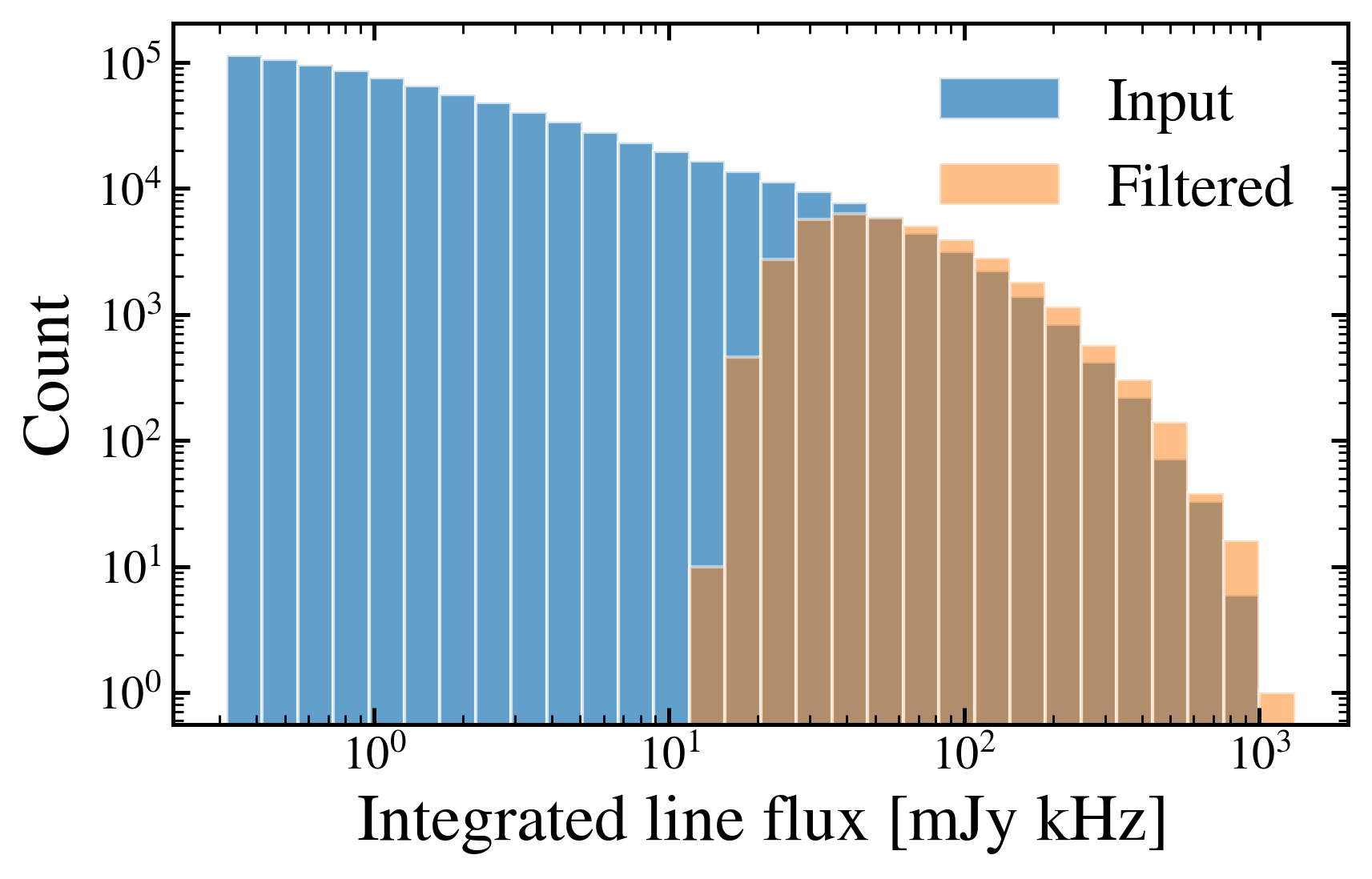}
\caption{Flux distribution of the input catalog and filtered \SOFIA\ detections.}\label{fig:input_filtered_galaxy}
\end{figure}

Based on tests using the development cube, additional filtering in a multi-dimensional parameter space was performed to substantially reduce the number of false positives, i.e. those detections without actual sources, and improve the reliability of the source catalog.
Specifically, we excluded detections that have \( n_{\rm pix} < 700 \), \( s < -0.00135 \times (n_{\rm pix} - 942) \), or \( f > 0.18 \times {\rm SNR} + 0.17 \), where \( n_{\rm pix} \) is the total number of spatial and spectral pixels contained within the three-dimensional source mask, \( s \) is the skewness of flux density values within the mask, \( f \) is the filling factor of the source mask within its 3D bounding box, and SNR is the integrated-flux signal-to-noise ratio of the detection.
These empirical relations were calibrated using tests on the development data cube and proved highly effective at rejecting spurious detections arising from random noise fluctuations.
The initial input catalog contains 2,392,131 sources.
The merged \SOFIA\ catalog includes 148,955 sources, which was reduced to 138,602 after the removal of duplicate entries.
Following unit conversion and the application of the above filtering criteria, the final catalog comprises 38,527 detections.

As a further check of the mock SKA-Mid catalog, the filtered \SOFIA\ detections were cross-matched with the corresponding true sources in the simulation.
The matching procedure followed the scoring method developed for the SKA Science Data Challenge\,2~\citep{Hartley2023}, which defines a quantitative framework for associating detected and true sources based on their spatial, spectral, and morphological properties.
The matching was carried out in the three-dimensional cube space, defined by right ascension (RA), declination (Dec), and central frequency.
A detection was considered a successful match if a true source lay within its spatial and spectral extent, as determined by the beam-convolved \HI\ size and linewidth of the detection.
Detections without any associated true source were classified as false positives.

According to the cross-matching, among the 38,527 detections, 36,872 of them are genuine sources.
This yields a global reliability of 95.7\%.
Figure~\ref{fig:input_filtered_galaxy} shows the flux distribution of the input catalog and the filtered \SOFIA\ detections.
The detection limit is about 6 Jy Hz.

\begin{figure}
    \centering
    \includegraphics[width=0.8\linewidth]{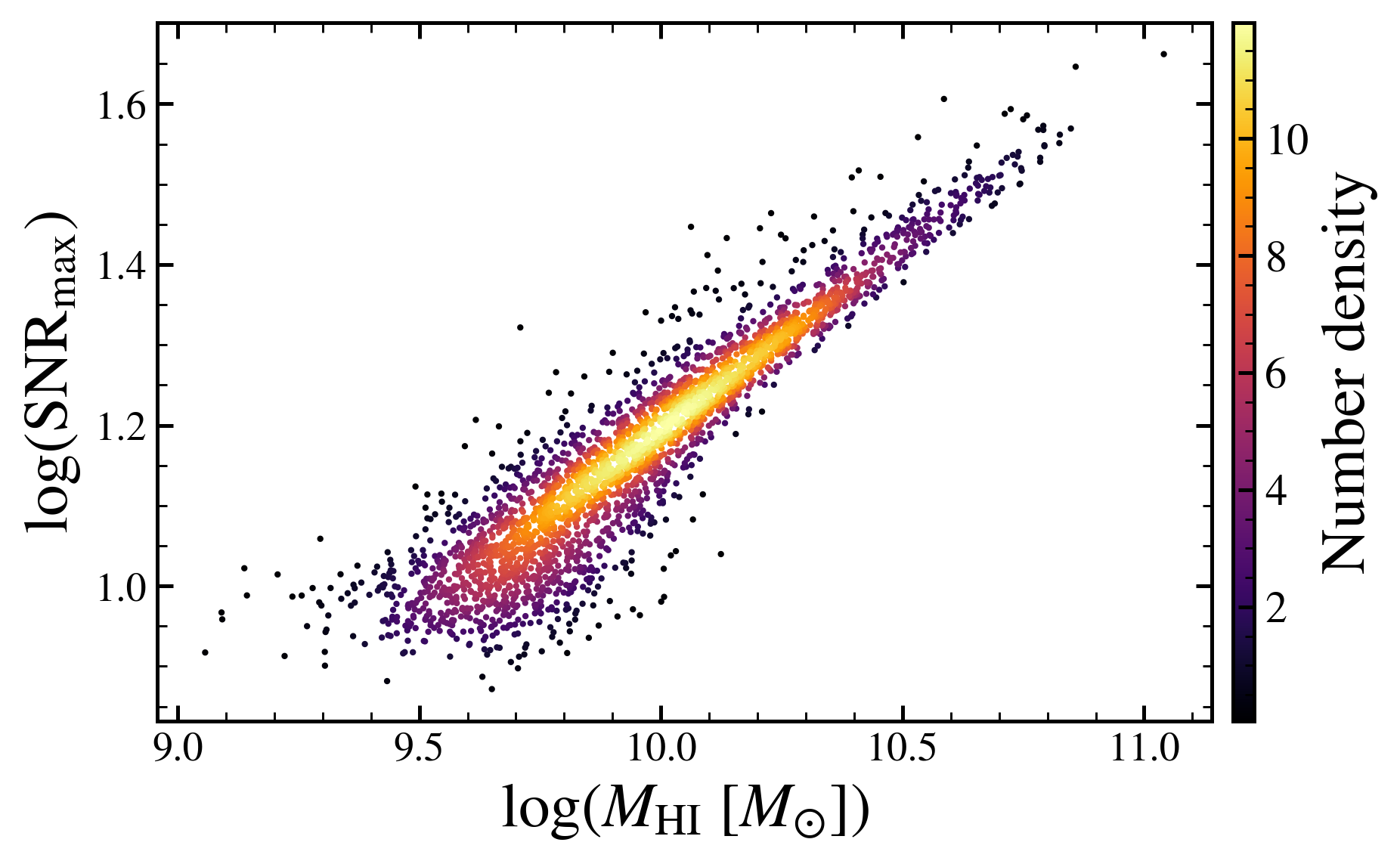}
    \caption{The ${\rm SNR}_{\max}$ as a function of \HI\ mass for SKA detectable galaxies at $0.34<z<0.36$.}\label{fig:snr_HI}
\end{figure}

For the matched source, the maximum SNR (${\rm SNR}_{\max}$) is derived in \SOFIA\ by integrating pixels in order of decreasing flux, starting from the brightest, and recalculating SNR until it peaks.
In Fig.~\ref{fig:snr_HI}, we show the ${\rm SNR}_{\max}$ as a function of the \HI\ mass for the detected galaxies at a thin redshift slice at $z\sim 0.35$.
The ${\rm SNR}_{\max}$ is basically proportional to the \HI\ density as one would expect, though at the low $M_{\rm HI}$ there is more spread, which may reflect the fact that other factors, such as the spatial and velocity distribution may also affect the detection.

\subsection{The CSS-OS galaxy catalog}\label{sec:gal_cat}

The CSST is equipped with seven photometric band filters (NUV, $u$, $g$, $r$, $i$, $z$, $y$), and three slitless spectrometers (GU, GV, GI) covering  $2,550$–$10{,}000~\text{\AA}$.
The CSS-OS will conduct a photometric imaging survey as well as a spectroscopic survey, covering a wide field of 17,500 ${\deg}^2$ in about 10 years. Below, we will consider the cross-correlation between the \HI\ survey and the CSS-OS spectroscopic survey.
\begin{figure}[ht!]
    \centering
    \includegraphics[width=0.8\linewidth]{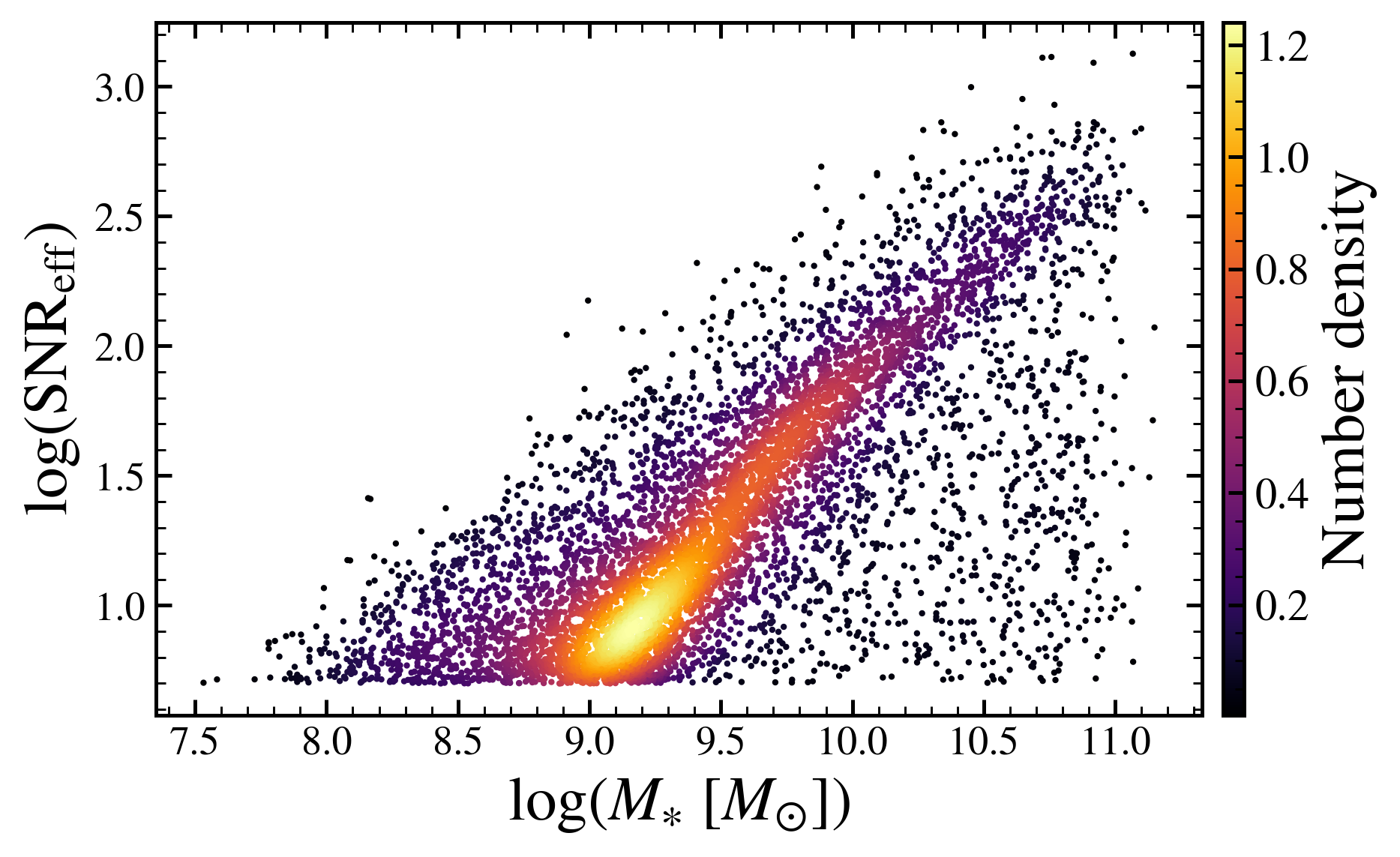}
    \includegraphics[width=0.8\linewidth]{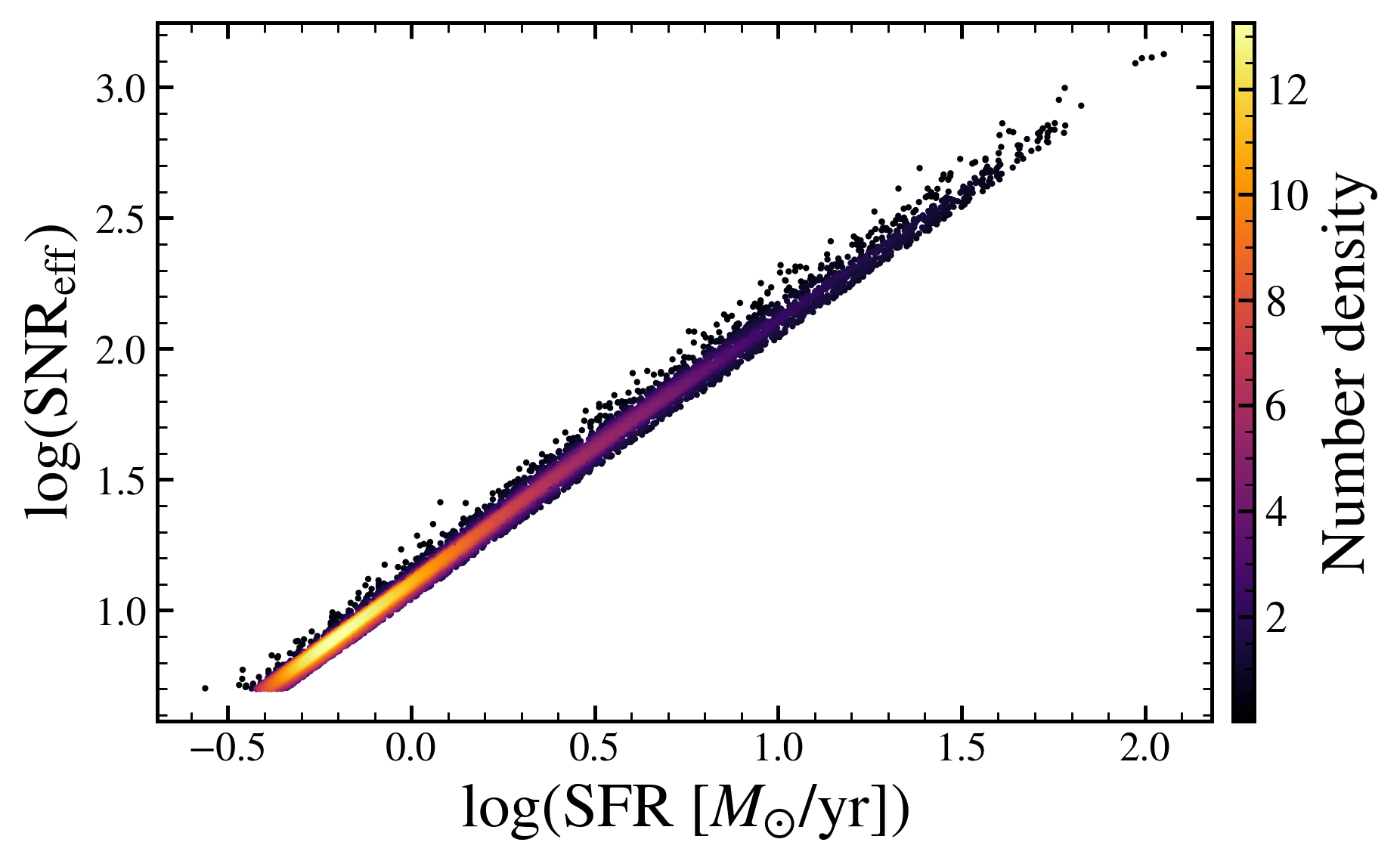}
    \caption{The scatter plot of the ${\rm SNR_{\rm eff}}$ of CSST spectroscopic galaxies as a function of stellar mass (top panel), and star formation rate (bottom panel) at $0.34<z<0.36$.
The color represents the number density in logarithmic scale.}\label{fig:snr-Ms-sfr}
\end{figure}

We use the strength of three strong emission-lines H\,$\alpha$, H\,$\beta$, and \OIII\ to access for whether a galaxy can be detected by CSST spectrometer.
For the slitless spectrometer, we model each galaxy as a point source for simplicity.
The SNR for each galaxy is calculated as~\citet{Deng2022}:
\begin{equation}
   \mathrm{SNR}=\frac{\left|\Delta C_s\right| t_{\exp} \sqrt{N_{\mathrm{obs}}}}{\sqrt{C_s t_{\exp}+N_{\mathrm{pix}}\left[\left(B_{\mathrm{sky}}+B_{\det}\right) t_{\exp}+R_{\mathrm{n}}^2\right]}}
\end{equation}
where $N_\txt{pix}=\Delta A/l_p^2$ is the number of pixels occupied by the aperture, 
$l_p=0.074\unit{arcsec}$ is the pixel size.
$B_{\det}=0.02\ e^-\txt{s^{-1}pixel^{-1}}$ 
is the detector dark current, $R_{\mathrm{n}}$ is the read noise, $t_{\exp}$ is the exposure time, and $N_\txt{obs}$ is the number of observations.
We set $t_{\exp}=150\unit{s}$ and $N_{\rm obs}=4$.
$B_\txt{sky}$ is the count rate from sky background per pixel:
\begin{eqnarray}
    B_\txt{sky} &=& A_\txt{eff}\int I_\txt{sky}(\nu)\,R_X(\nu)\,l_p^2\, \frac{d\nu}{h\nu},
\end{eqnarray}
where $A_\txt{eff}=3.14 \unit{m^2}$ is the effective area for CSST, and $I_\txt{sky}(\nu)$ is the average sky background, including earthshine and zodiacal light~\citep{Ubeda2011}.
$C_s$ is the count rate from the galaxy, $\Delta C_s$ is the signal contrast.
For the spectrometer and the spectral line $i$, 
\begin{eqnarray}
    \Delta C_s^i &=& A_\txt{eff}R_X\left(\frac{\nu_i}{z+1}\right)\frac{\Delta F_\txt{line}^i}{h\nu_i/(z+1)}, \\
    C_s^i &=& \Delta C_s^i + A_\txt{eff}\int_{\nu_o^i-\nu_R/2}^{\nu_o^i+\nu_R/2} R_X F_\txt{cont}\frac{1}{h\nu}d\nu,
\end{eqnarray}
where $\nu_o^i$ is the observed frequency for line $i$ and $\nu_R$ is the spectral resolution of CSST spectrometer.
$\Delta F_\txt{line}$ and $F_\txt{cont}$ are the flux contrast and continuum detected by the CSST spectrometer, given by:
\begin{align}
\label{eq:dfline}
\Delta F_\txt{line}^i &= \frac{1}{4\pi r_L^2}\int_{\nu_i-\frac{\Delta\nu}{2}}^{\nu_i+\frac{\Delta\nu}{2}}(L_\txt{rest}(\nu) - L_\txt{rest}^\txt{cont}(\nu))d\nu,\\
\label{eq:fcont}
F_\txt{cont}(\nu) &= \frac{z+1}{4\pi r_L^2}L_\txt{rest}^\txt{cont}\,\left((z+1)(1+\frac{v_\txt{los}}{c})\nu\right),
\end{align}
where $\nu_i$ is the rest-frame wavelength for line $i$, $\Delta\nu$ is the size of the window, and $L_\txt{rest}^\txt{cont}$ is the continuum luminosity.
We assume that the emission features are narrow enough to be fully covered by a single CSST spectrometer resolution unit.
We define an effective signal-to-noise ratio for each line as
\begin{equation}
    {\rm SNR}_{\rm eff} \equiv \frac{\Delta C_s}{C_{\lim}} .
\end{equation}
A galaxy is considered detected if at least one emission-line satisfies ${\rm SNR_{\rm eff}}>5$.
Figure~\ref{fig:snr-Ms-sfr} shows the CSST observation ${\rm SNR_{\rm eff}}$ as a function of the stellar mass (top panel), and as a function of the star formation rate (SFR) (bottom panel), at a thin redshift slice.
There is a trend of increasing ${\rm SNR}_{\rm eff}$ with increasing stellar mass, but the scatter is quite large, and there are many massive galaxies with low ${\rm SNR}_{\rm eff}$.
The spectroscopic ${\rm SNR}_{\rm eff}$ is tightly correlated with the SFR, for the emission-lines are primarily determined by the SFR.

\section{Results}\label{sec:result}

We now present the results from the simulation.
First, for the detected sample, we show the halo mass, \HI\ mass, stellar mass, and star formation rate, to reveal their relations.
We then plot their number density evolution over redshift.
Next, we show the simulated Baryonic Tully-Fisher relationship for the \HI\ detected sample, which connects the optical observation to the radio observation.
Given that the optical observation of the CSST is deeper than the SKA \HI\ observation, more information can be extracted by stacking the \HI\ observation results for some of the optically detected galaxies.
Finally we quantify the clustering strength of the \HI\ selected sample by measuring the \HI-optical galaxy cross-correlation function.

\subsection{The galaxy samples}

\begin{figure}[ht!]
\centering
\includegraphics[width=0.8\linewidth]{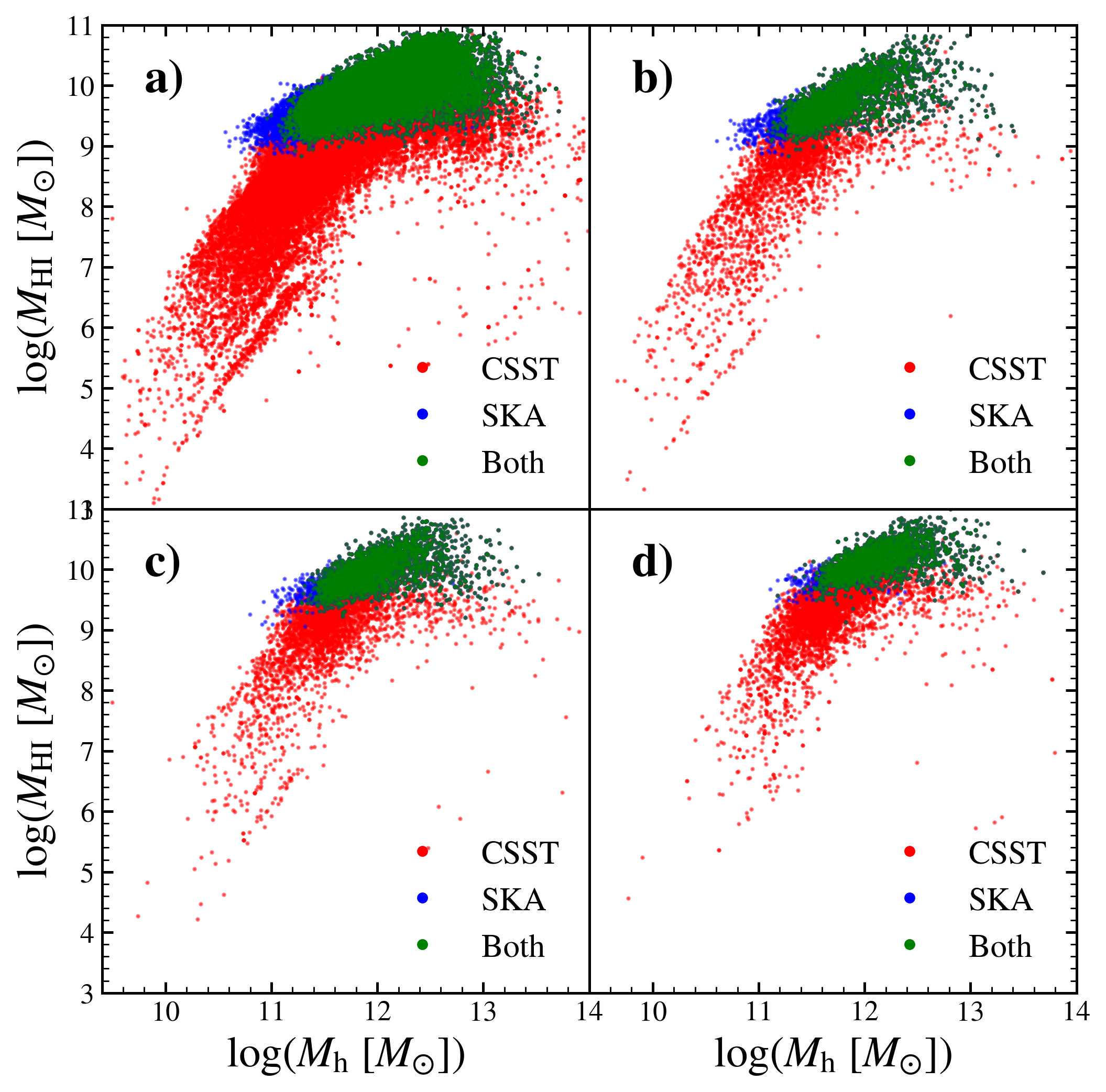}
\caption{A scatter plot of the \HI\ mass and halo mass of detected galaxies.
Panel \texttt{a} corresponds to the whole redshift range ($z = 0.235 \sim 0.495$), panel \texttt{b}, \texttt{c} and \texttt{d} correspond to three redshift slices at  $0.24$--$0.26$, $0.34$--$0.36$, $0.44$--$0.46$ respectively.}\label{fig:MHi-Mh}
\end{figure}

To highlight the difference between CSST-detected and SKA-detected samples, we make a scatter plot of the halo mass and \HI\ mass of the galaxies detected by the SKA \HI\ survey (blue) and the CSST spectroscopic survey red, and those detected by both (green) in Figure~\ref{fig:MHi-Mh}. Here we show the result for all galaxies in the lightcone in panel (a), while panels (b)–(d) display the distributions in three redshift slices centered at $z=0.25$, $0.35$, and $0.45$, respectively, each with a width of $\Delta z = 0.02$.
The CSST spectroscopic survey is more sensitive than the SKA \HI\ survey, and therefore enables more detections for galaxies with low \HI\ masses and low halo masses. Most of the galaxies detectable by the SKA \HI\ survey are also detectable by the CSST spectroscopic survey, thus the blue and green dots are largely overlapping with each other.
However, there are some SKA-detectable galaxies with high \HI\ masses but relatively low halo masses which are not detectable with CSST survey.
At higher redshifts, the detectable galaxies are generally fewer in all surveys.
\begin{figure}[h]
\centering
\includegraphics[width=0.9\linewidth]{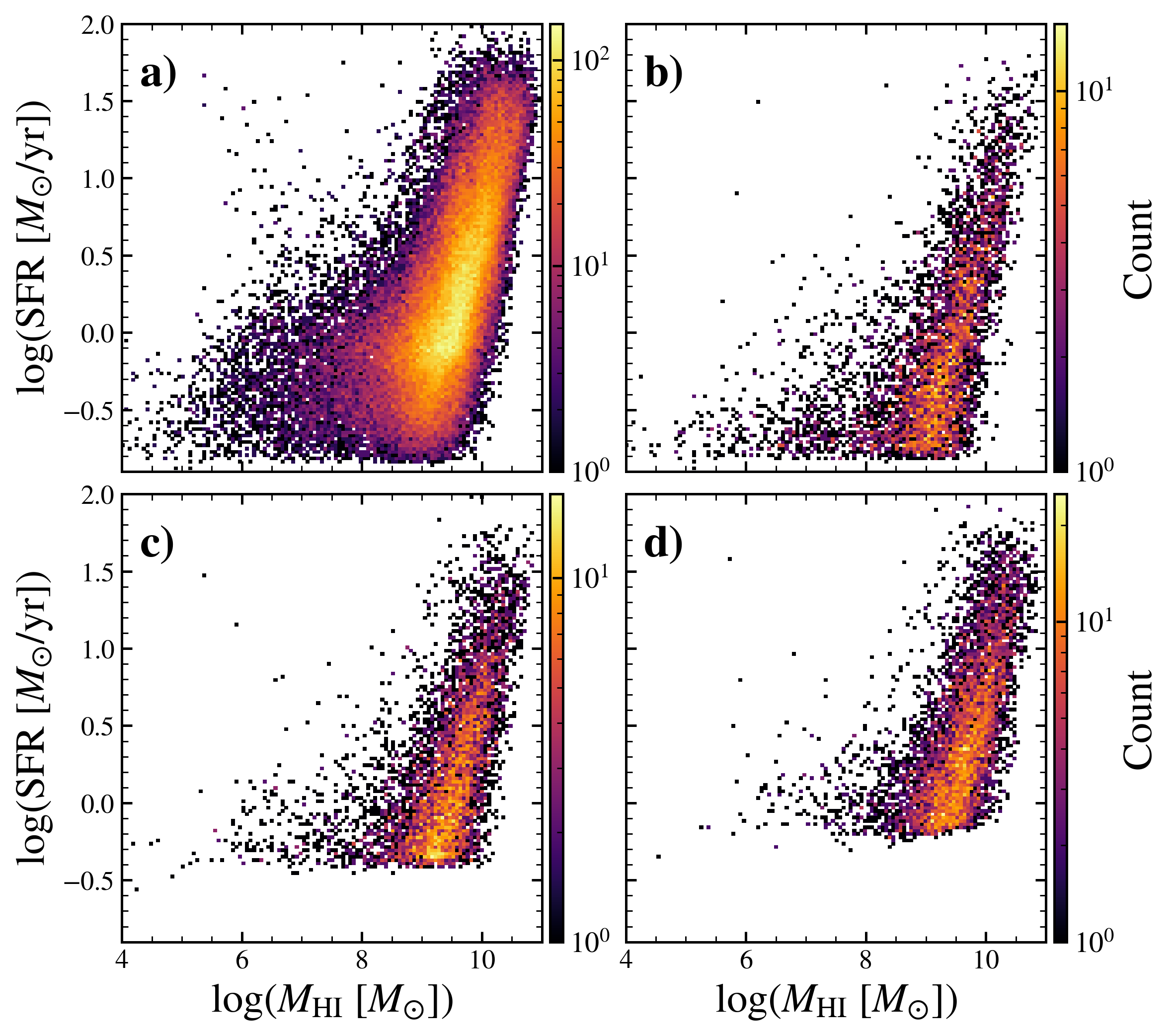}
\caption{The relations between star formation rate (SFR) and \HI\ mass.
Panel \texttt{a} corresponds to the whole redshift range ($z = 0.235 \sim 0.495$), panel \texttt{b}, \texttt{c} and \texttt{d} correspond to three redshift slices centered at $0.25$, $0.35$, and $0.45$ respectively.}\label{fig:sfr-mhi-csst}
\end{figure}

We have shown earlier that the emission-line SNR is proportional to the SFR, while the \HI\ SNR is proportional to \HI\ mass.
To better understand the galaxies which are detectable with both the SKA \HI\ survey and the CSST spectroscopic survey, we make a scatter plot of the SFR and \HI\ mass of such galaxies in Figure~\ref{fig:sfr-mhi-csst}.
Similar to Fig.~\ref{fig:MHi-Mh}, we make the plot in four sub-panels: (a) for all of the lightcone, and (b), (c), (d) for redshift redshift slices centered at $z=0.25$, $0.35$, and $0.45$, respectively.
As can be seen from the figure, the SFR and \HI\ mass is positively correlated, especially if one looks at a single redshift, though there are some scatters, with some galaxies which have relatively low \HI\ mass still having high SFR, which corresponds to those galaxies which are not detectable with SKA \HI\ survey but can be detected by the CSST spectroscopic surveys.
\begin{figure}
\centering
\includegraphics[width=0.9\linewidth]{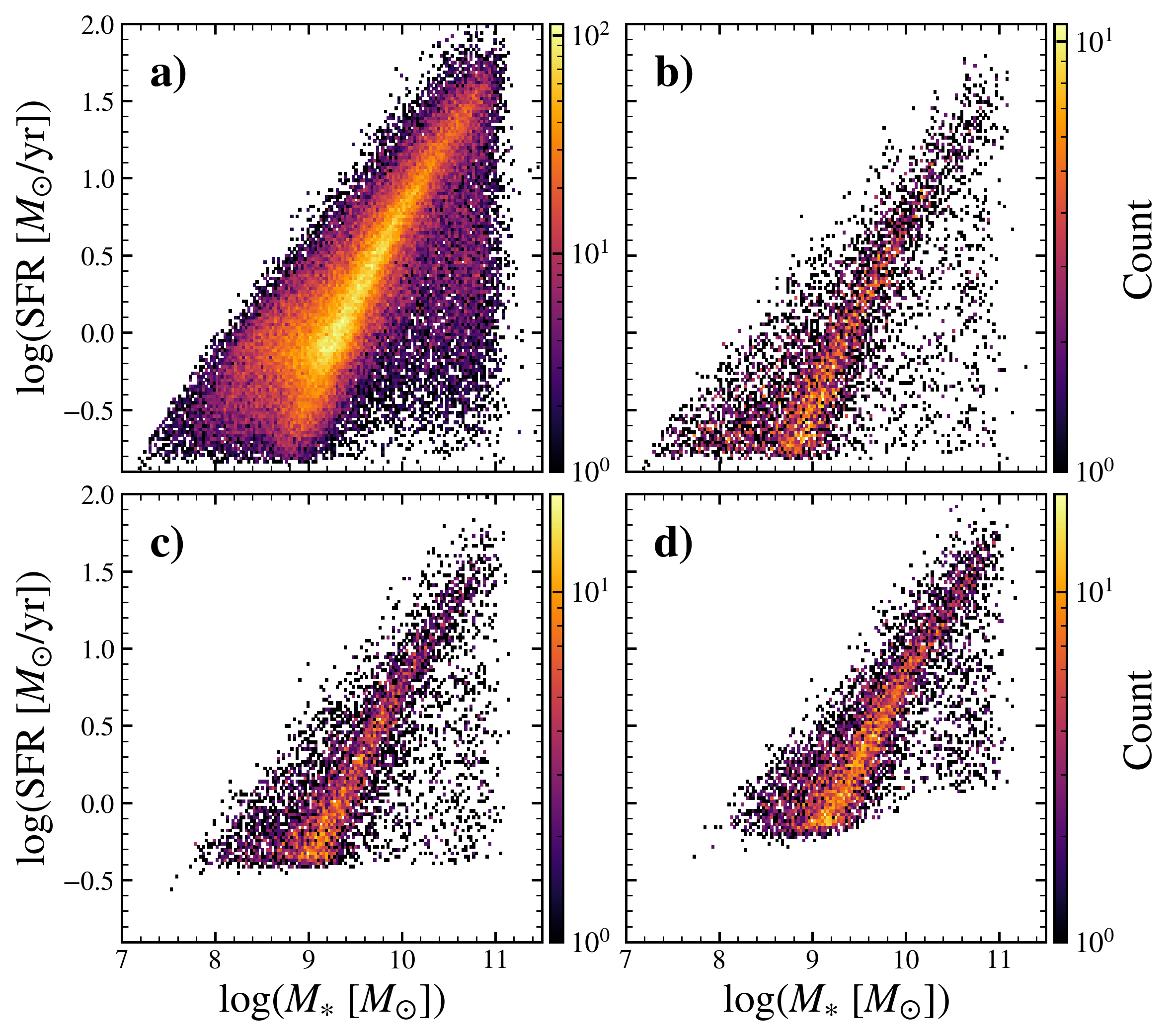}
\caption{The relations between star formation rate (SFR) and stellar mass.
Panel \texttt{a} corresponds to the whole redshift range ($z = 0.235 \sim 0.495$), panel \texttt{b}, \texttt{c} and \texttt{d} correspond to three redshift slices centered at $0.25$, $0.35$, and $0.45$ respectively.}\label{fig:sfr-ms-csst}
\end{figure}

Note that the CSST spectroscopic survey detects galaxies with high SFR, which is different from the photographic survey, which detects galaxies that are bright in optical continuum, and follows more closely to the total stellar mass.
In Fig.~\ref{fig:sfr-ms-csst}, we plot the SFR vs. $M_{\ast}$, for the whole lightcone (sub-panel a) and the same three redshift slices.
The SFR is also positively correlated with the $M_{\ast}$, but there are large spreads, so the galaxies bright in photographic surveys are not always detectable in its emission-lines, and vice versa.

\subsection{Number densities}

\begin{figure}[ht!]
\centering
\includegraphics[width=0.8\linewidth]{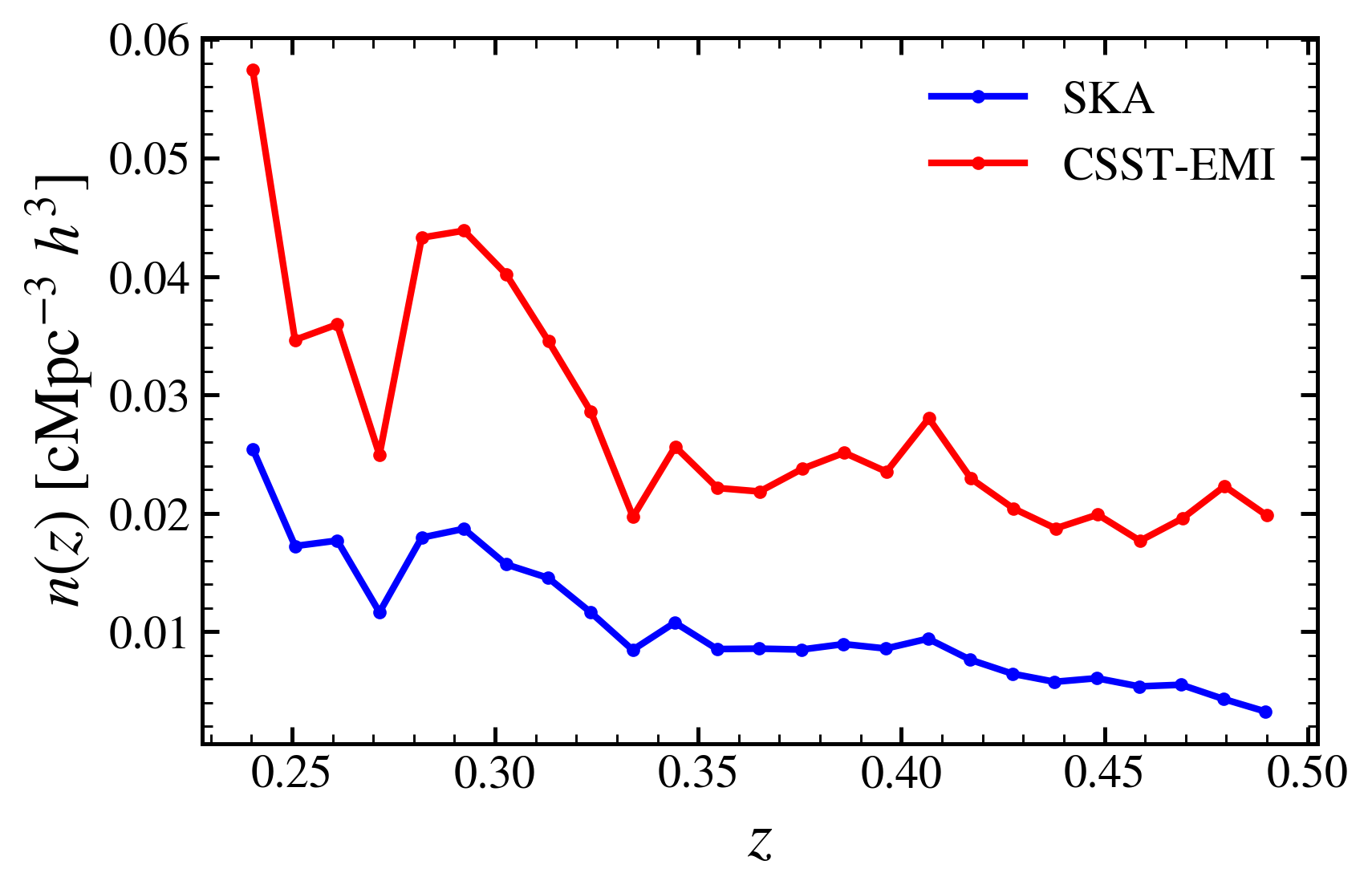}
\caption{The number density of simulated galaxies detected by the SKA-Mid survey and CSST survey as a function of redshift.\label{fig:galdens_ska}}
\end{figure}

\begin{figure}[ht!]
    \centering
    \includegraphics[width=0.8\linewidth]{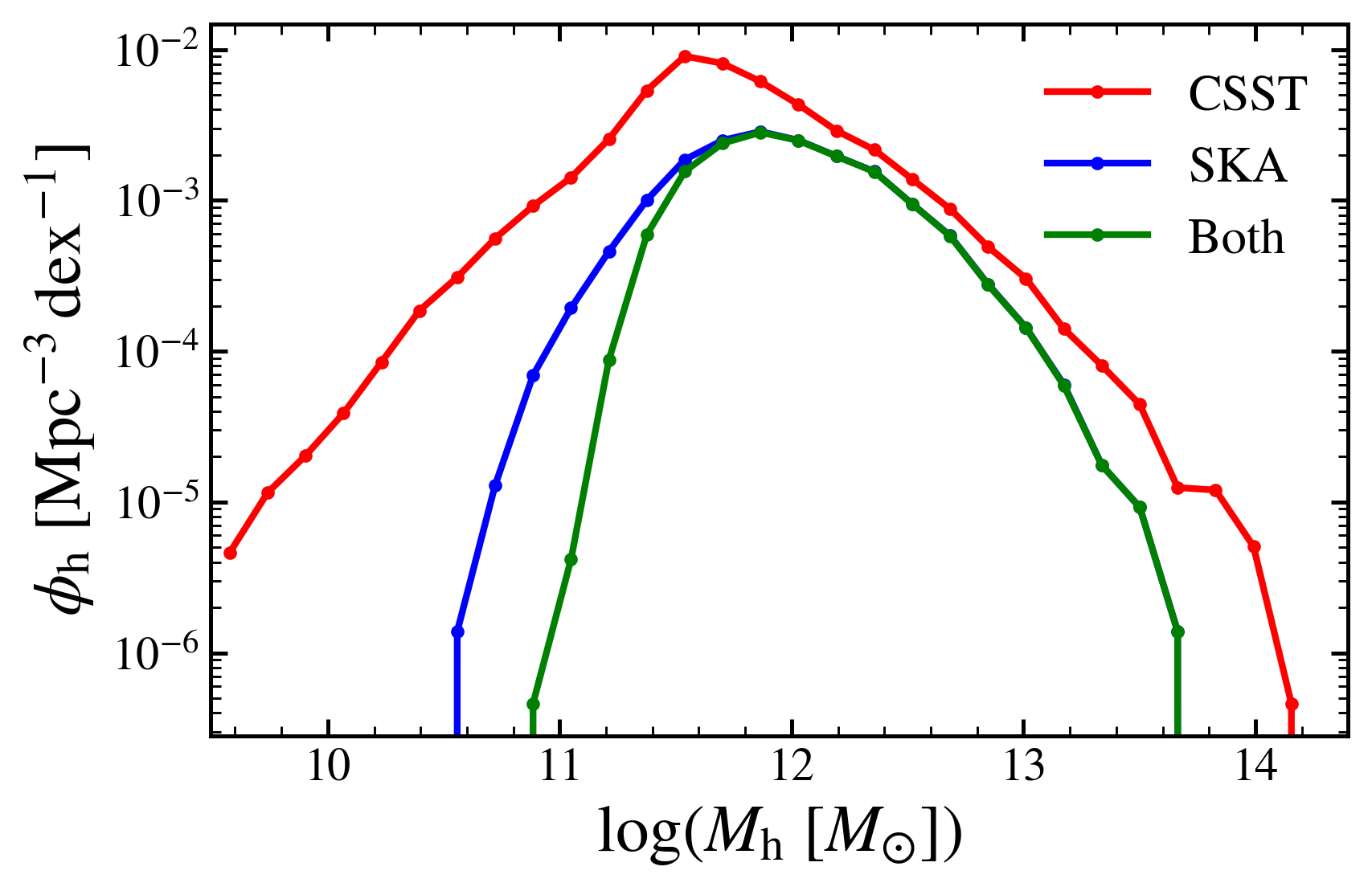}
    \caption{The number density of the galaxies as a function of halo masses.}\label{fig:galden-Mh}
\end{figure}

In Figure~\ref{fig:galdens_ska} we show the number densities of the galaxies detectable with SKA-Mid \HI\ survey and the CSST spectroscopy survey.
The overall trend are similar, with the number density decreasing with increasing redshift.
The number density of the CSST-detected galaxies is higher than the SKA-detected \HI\ galaxies, but this of course depends on the total observing time for both surveys.
We can see the change of the number density over redshift is not monotonic but varies with fluctuations due to sample variance, since the density is sampled in slices smaller than the whole simulation box.
The fluctuations in the SKA and CSST galaxies are highly similar to each other, which is expected, since both trace the same underlying density fluctuations in the simulation box.
In Figure~\ref{fig:galden-Mh}, we show the mass function of the galaxies detected by the CSST spectroscopic survey and the SKA \HI\ survey in the lightcone as a function of their halo mass.
These results are consistent with what have seen in Figure~\ref{fig:MHi-Mh}.
As note above, the CSST survey is more sensitive, thus it will detect more galaxies, which span a wider range of halo mass.
The number density of SKA galaxies is smaller, and it drops sharply below $M_h = 10^{11} \Msun$.
At the higher halo mass end, almost all SKA-detectable galaxies are also detectable by the CSST, thus the galaxies detectable by both have almost the same density as the SKA detectable galaxies.
Nevertheless, some CSST-detectable galaxies in massive halos with $M_h \sim 10^{14}\Msun$ are not detectable by the SKA, probably because the \HI\ mass of these galaxies is not high.

\subsection{Baryonic Tully–Fisher relation}

The Tully-Fisher relation (TFR;~\cite{Tully1977}) is one of the most important empirical scaling relations in extragalactic astronomy, linking the intrinsic luminosity of a spiral galaxy to its rotation velocity.
Physically, it reflects the connection between the depth of a galaxy’s potential well, dominated by its dark matter halo, and the total stellar content assembled through baryonic processes.
As the CSST and SKA will produce high quality observational data at higher redshifts in the optical and \HI\ band respectively, we also check this relation using the simulated sample.

Over the years, the TF relation has been extensively used as a distance indicator, as well as a probe of galaxy formation and evolution.
However, it varies with galaxy properties, including colors and ages.~\citet{McGaugh2000} proposed that the baryonic Tully-Fisher relation (BTFR), which replaces luminosity or stellar mass with the total baryonic mass (stars plus cold gas), may produce a better correlation than the original TFR, and is possibly a more fundamental relation~\citep{McGaugh2000, Lelli2016} from the perspective of the modified Newtonian Dynamics (MOND) scenario. 
In the present study, the relation is however derived from a $\Lambda$CDM model, and it is a result of the general scaling relations in that model.
\begin{figure}
\centering
\includegraphics[width=0.9\linewidth]{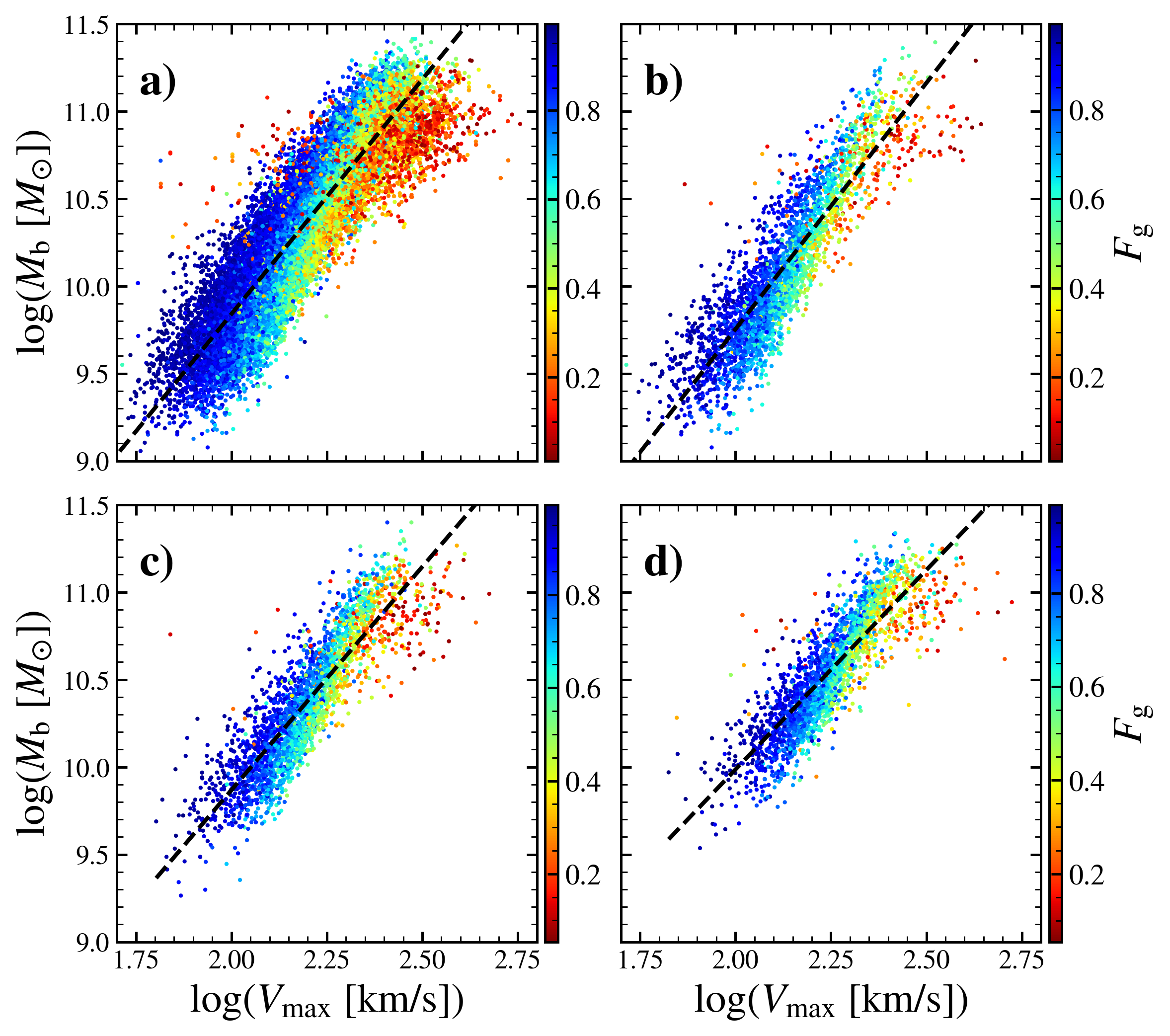}
\caption{The BTFR for the \HI-detected sources.
Panel \texttt{a} corresponds to the whole redshift range $0.235 \sim 0.495$, panel \texttt{b}, \texttt{c} and \texttt{d} correspond to three redshift slices centered at $0.25$, $0.35$, and $0.45$ respectively.}\label{fig:BTFR-SKA}
\end{figure}

Figure~\ref{fig:BTFR-SKA} shows the BTFR for the SKA \HI-detected sources.
We show the results for the whole lightcone and for the redshift slices centered at $z=0.25, 0.35, 0.45$ respectively.
The sources are colored by the cold gas fraction of the baryons, $F_g \equiv \log (M_g/M_b)$, where $M_b = M_g + M_{\ast}$ is the baryonic mass and $M_g$ is the cold gas mass.
The baryonic mass is well correlated with the maximum rotation velocity, even though the cold gas fraction varies significantly over the range.

\subsection{Stacking analysis}

To probe the average \HI\ content of galaxies below the individual detection threshold of SKA, we perform a stacking analysis using galaxies selected from the CSST spectroscopic catalog.
This approach allows us to statistically recover the mean \HI\ signal by co-adding the emission from galaxies with known positions and redshifts.
The stacking sample consists of galaxies detected via CSST emission-lines, which provide accurate sky coordinates and redshifts, but are not necessarily individually detected in \HI\ by SKA.
The galaxies are divided into bins of redshift and stellar mass, and average quantities are computed within each bin.

For each galaxy, a subcube (``mini cube'') is extracted from the full \HI\ data cube based on its sky position and intrinsic extent in both angular and frequency directions.
The size of the subcube is determined in a data-driven manner.
For each redshift bin, we estimate the typical spatial extent and velocity width of galaxies from the distributions of intrinsic \HI\ sizes and line widths in the mock catalogue.
We adopt the 95th percentile of these distributions to define the angular diameter and velocity width of the subcube, ensuring that the majority of the \HI\ emission is enclosed while avoiding excessive noise contamination.
We have also examined the impact of the subcube size on the stacked \HI\ mass.
We find that adopting a smaller cube size may lead to a loss of flux from extended emission, while a significantly larger cube size increases the noise contribution without substantially changing the recovered signal.
Therefore, our choice represents a balance between signal completeness and noise control, and does not significantly bias the stacked \HI\ mass.

\begin{figure}[th]
\centering
\includegraphics[width=0.6\linewidth]{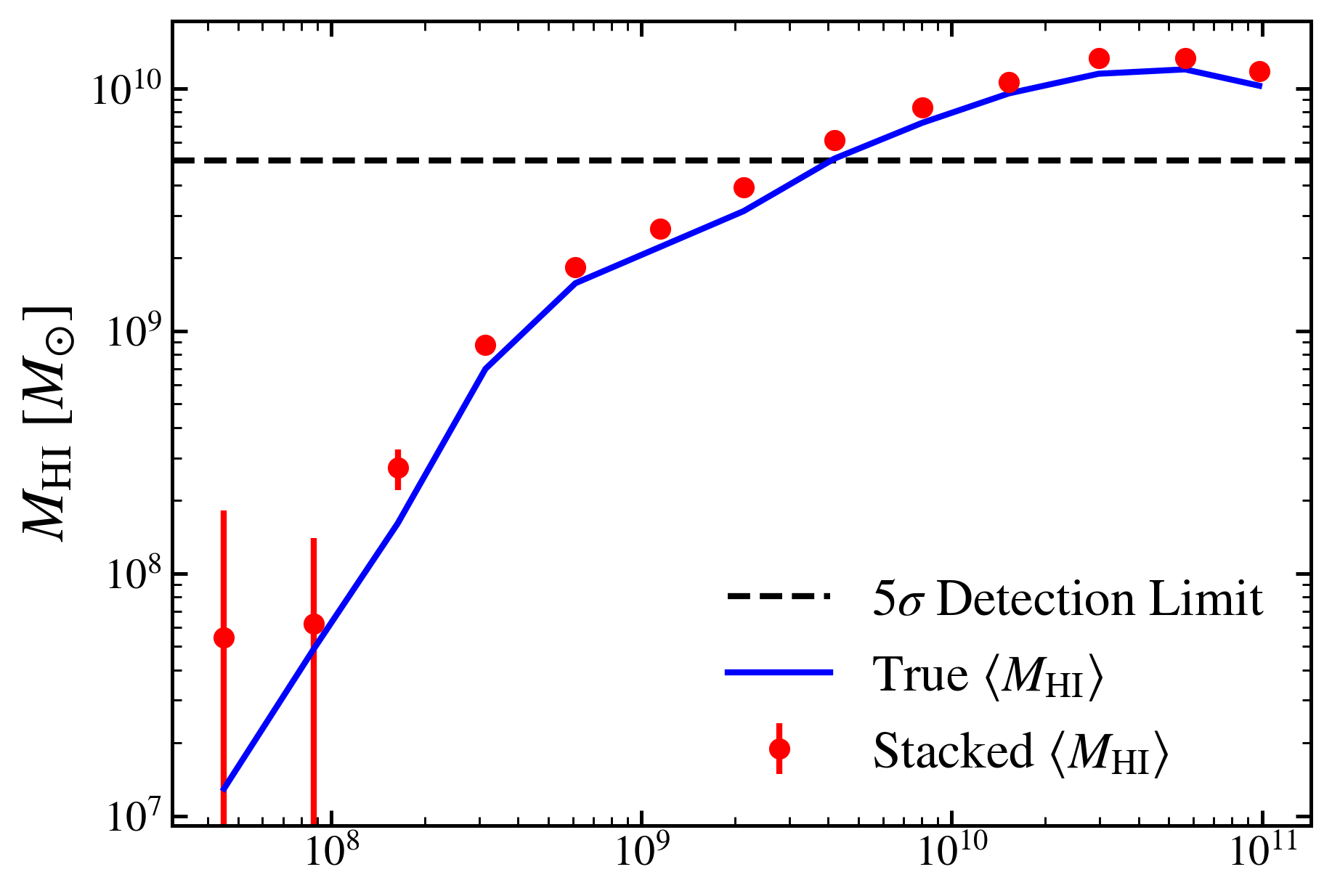}
\includegraphics[width=0.6\linewidth]{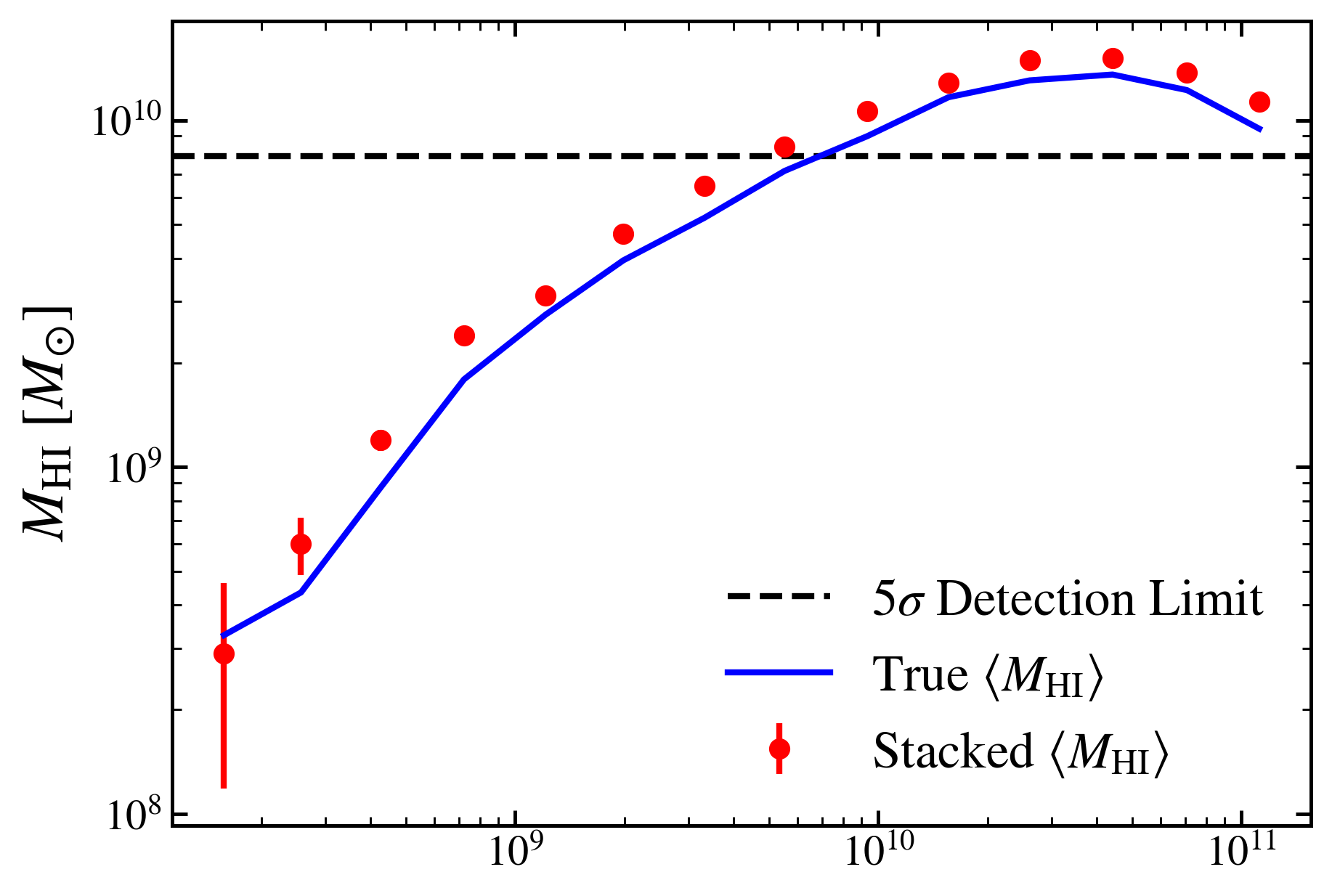}
\includegraphics[width=0.6\linewidth]{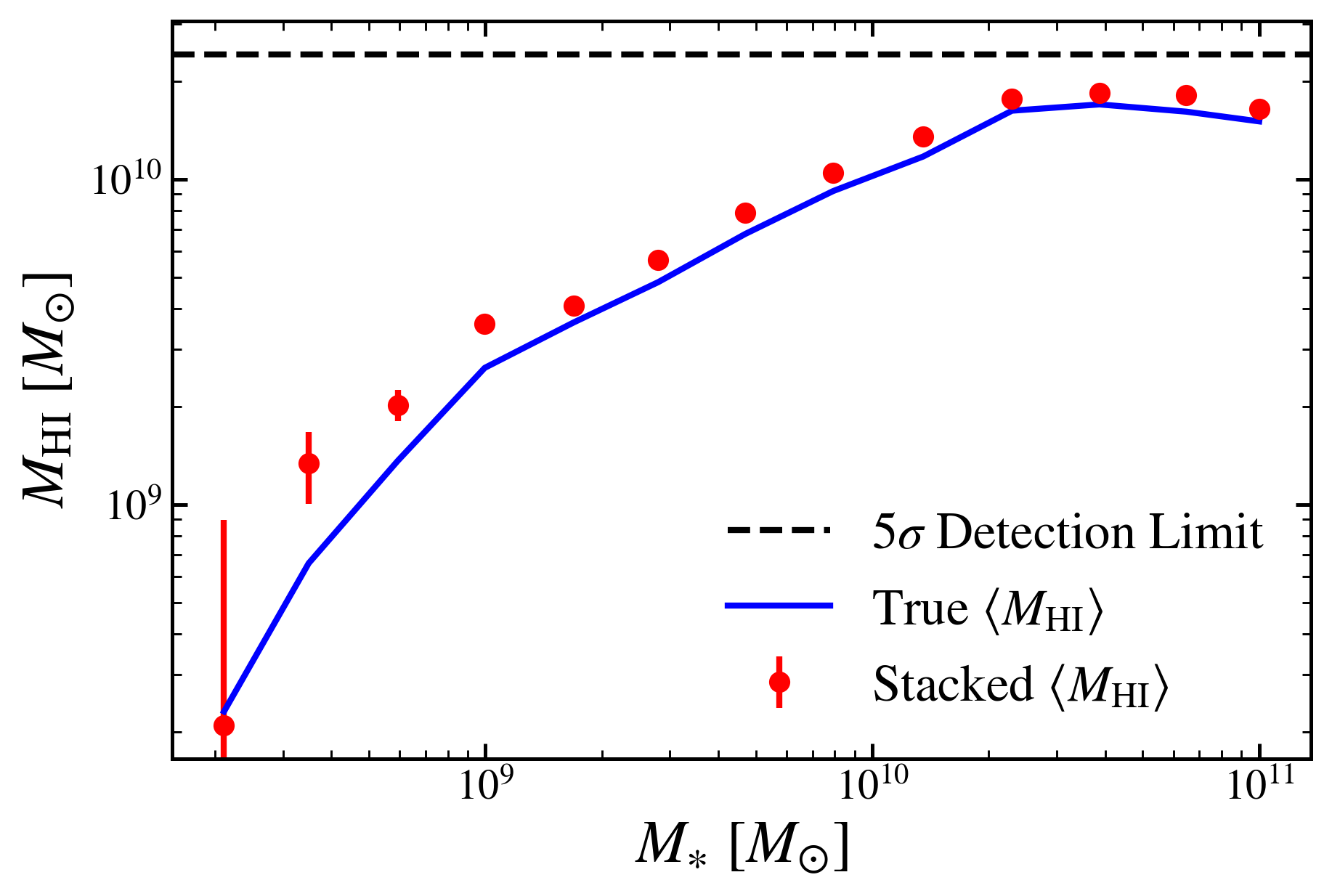}
\caption{Comparison between the stacked (points) and intrinsic (curve) \HI\ masses as a function of stellar mass in three redshift bins (top panel: $z\approx0.28$; middle panel: $z\approx0.38$; bottom panel: $z\approx0.48$).
The vertical error bars indicate the statistical uncertainty estimated from noise realizations.
The horizontal dashed line denotes the $5\sigma$ detection limit for individual galaxies.}\label{fig:stacking_Ms_MHi}
\end{figure}

The integrated flux is computed by summing over all voxels within the subcube:
\begin{equation}
S = \sum_{i}\, \sum_{(x,y)} I_i(x,y)\, \Delta v_i,
\end{equation}
where $I_i(x,y)$ is the intensity in Jy/pixel and $\Delta v_i$ is the velocity width of channel $i$.
This procedure ensures that the full \HI\ emission associated with each galaxy is included, without imposing an idealized aperture, thereby minimizing aperture-induced biases.
The integrated flux is converted to \HI\ mass using
\begin{equation}\label{eq:flux2mass}
\frac{M_{\mathrm{\HI}}}{\mathrm{M}_{\odot}} = 2.36 \times 10^5 
{\left(\frac{D_{\mathrm{L}}}{\mathrm{Mpc}}\right)}^2
\,\frac{S_{\nu}}{\mathrm{Jy}\,\mathrm{km\,s^{-1}}} {(1+z)}^{-2},
\end{equation}
where $S_{\nu}$ is the integrated flux in Jy km\,s$^{-1}$ and $D_{\rm L}$ is the luminosity distance in \Mpc.
For a given bin containing $N$ galaxies, the stacked \HI\ mass is computed as the mean:
\begin{equation}
\langle M_{\mathrm{\HI}} \rangle = \frac{1}{N} \sum_{j=1}^{N} M_{{\mathrm{\HI}},\,j}
\end{equation}
The stacking is performed on data cubes that include both signal and noise, providing a realistic estimate of observational recovery.
The uncertainty of the stacked \HI\ mass is dominated by instrumental noise.
To estimate this uncertainty, we perform integrations over randomly placed cylindrical apertures in a noise-only cube.
The spatial radius and velocity width of the cylinders are chosen to match the typical \HI\ extent of galaxies within each redshift bin.
This procedure yields a distribution of integrated fluxes arising purely from noise, from which we compute the standard deviation $\sigma_{\rm flux}$.
The corresponding \HI\ mass uncertainty, $\sigma_{M_{\rm HI}}$, is obtained by converting the flux dispersion using the same relation as Eq.~(\ref{eq:flux2mass}).
For a stack of $N$ galaxies, the uncertainty scales as
\begin{equation}
\sigma_{\rm stack} = \frac{\sigma_{M_{\rm HI}}}{\sqrt{N}},
\end{equation}
assuming that the noise contributions from different galaxies are uncorrelated.
To assess the accuracy of the stacking measurement, we compare the recovered \HI\ mass with the intrinsic values from the simulation.
The true mean \HI\ mass in each bin is computed directly from the underlying galaxy catalog, without any observational noise or measurement effects.
This comparison allows us to quantify potential biases and assess the reliability of the stacking technique in recovering the underlying \HI\ properties of the galaxy population.
Figure~\ref{fig:stacking_Ms_MHi} shows the comparison between the stacked \HI\ masses and the true mean values as a function of stellar mass in three representative redshift bins ($z\approx0.28,\ 0.38,\ 0.48$ from top to bottom panel, respectively).
The dashed line indicates the $5\sigma$ detection limit for individual galaxies.
At the low stellar mass end ($M_* \lesssim 10^9\,M_{\odot}$), most galaxies fall below the individual detection threshold.
Nevertheless, the stacking analysis is able to recover a statistically significant \HI\ signal.
The stacked $\langle M_{\rm HI} \rangle$ values broadly follow the trend of the true relation, demonstrating that stacking effectively probes the average \HI\ content of galaxies that are otherwise undetectable individually.
However, the uncertainties become significantly larger in the lowest stellar mass bins.
This reflects the decreasing SNR, as the intrinsic \HI\ emission becomes comparable to the noise level.
As a result, the stacking measurements in this regime are less well constrained.

At intermediate and high stellar masses ($M_* \gtrsim 10^9\,M_{\odot}$), the agreement between the stacked and true \HI\ masses is significantly improved.
The stacked measurements closely trace the intrinsic relation, indicating that the stacking procedure provides a reliable estimate of the mean \HI\ content when sufficient signal is present.
A mild systematic excess of the stacked values over the true relation is observed at the high-mass end.
This trend is likely attributable to selection effects in the CSST emission-line sample, which preferentially selects gas-rich galaxies, as well as potential noise boosting in the stacking process.

\subsection{Clustering strength of \HIt\ selected sample}

After obtaining galaxy catalogs from SKA and CSST, we derive the clustering strength of the \HI\ galaxies, i.e., the linear bias of the SKA catalog with respect to the underlying dark matter field, in three redshift bins centered at $z=0.28$, $0.36$, and $0.44$, each with a comoving depth of $200\,{\rm Mpc}/h$. 

\begin{figure}[t]
    \centering
    \includegraphics[width=0.8\linewidth]{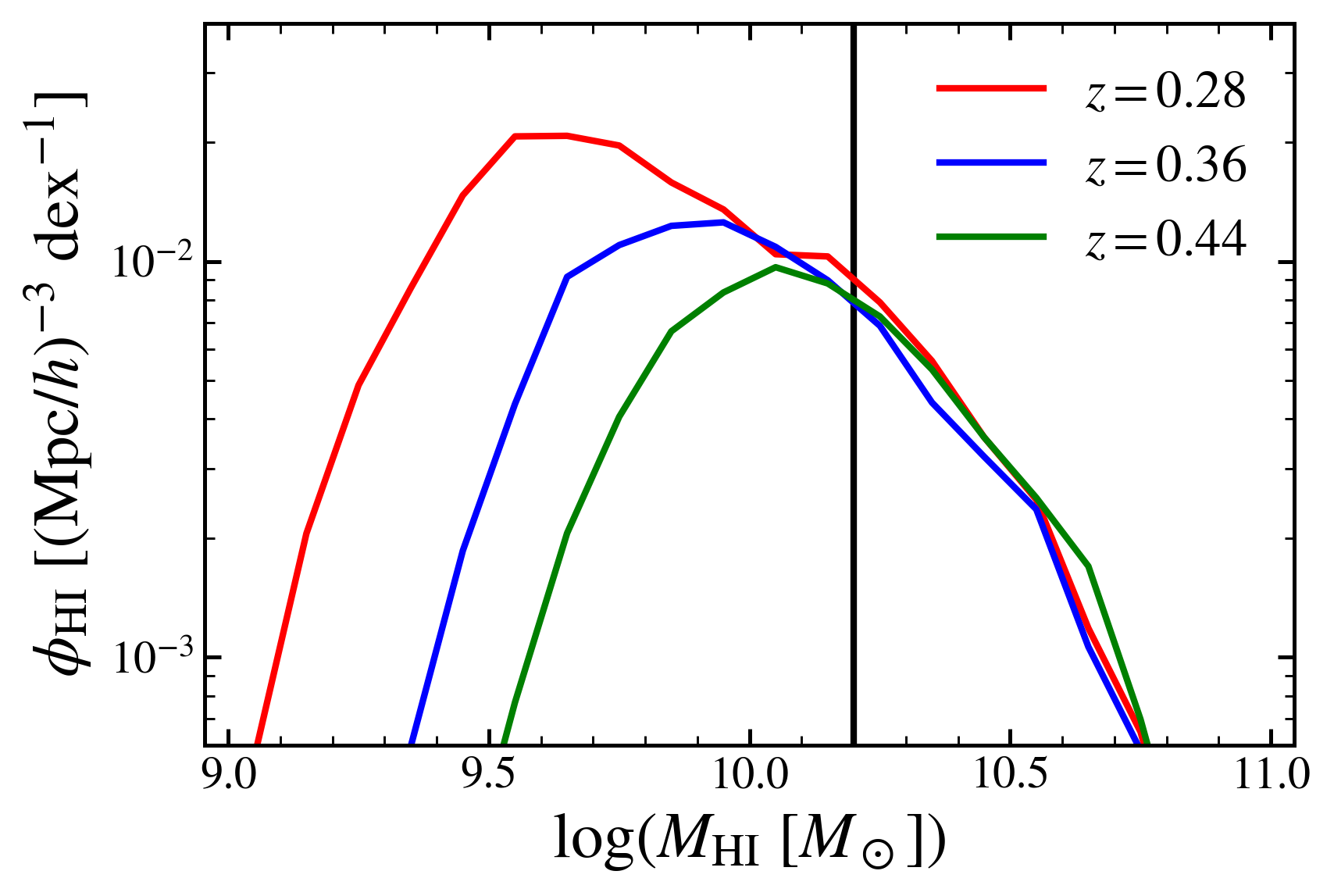}
    \includegraphics[width=0.8\linewidth]{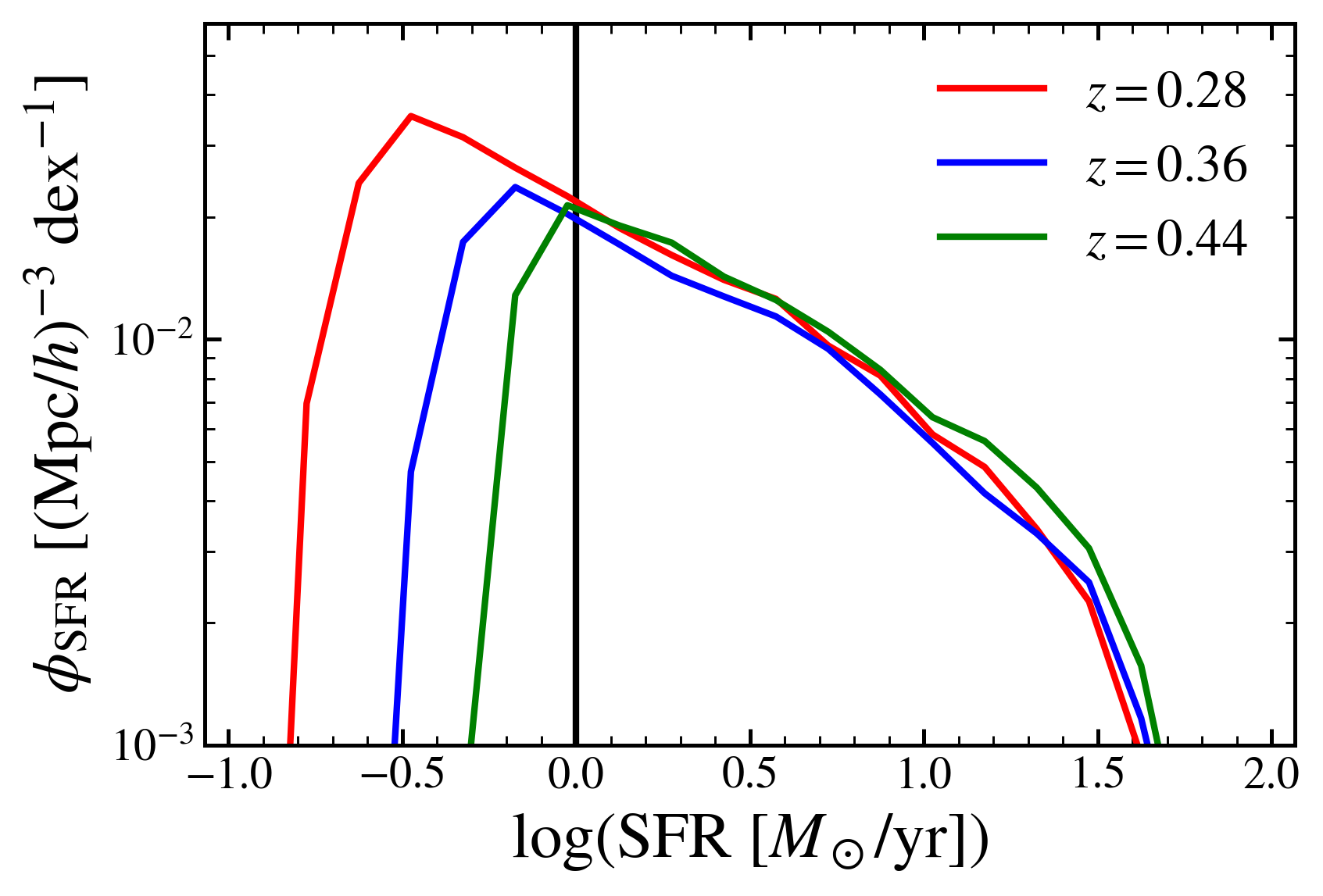}
    \caption{The \HI\ mass function of SKA catalog (upper panel) and SFR distribution of the CSST catalog (lower panel) for different redshift bins.
The vertical solid lines mark $M_{\rm HI}=10^{10.2}\,M_{\odot}$ (upper panel) and ${\rm SFR}=1\,M_\odot/{\rm yr}$ (lower panel), which are the selection criteria for the corresponding catalogs.}\label{fig:selection_crit}
\end{figure}

Leveraging the high number density of the CSST catalog, we compute the cross-correlation between SKA- and CSST-detected galaxies to improve the SNR.
To mitigate redshift-dependent selection effects, we apply cuts of $M_{\rm HI}>10^{10.2}\,M_{\odot}$ for the SKA sample and $\mathrm{SFR}>1\,M_\odot/\mathrm{yr}$ for the CSST sample as shown in Figure~\ref{fig:selection_crit}, yielding approximately uniform number densities across the redshift bins.
We have also verified that the stellar mass and halo mass functions are consistent across bins, indicating that these selections effectively minimize redshift-dependent biases.

Then we calculate the cross power spectrum between the \HI-selected catalog and the optical-selected catalog.
We model the cross power spectrum as
\begin{equation}
    P_X(k) = b_{\rm opt} b_{\rm HI} P_{\rm lin}(k) + P_X^{\rm SN}
\end{equation}
where $b_{\rm opt}$ and $b_{\rm HI}$ are the linear biases of the optical-selected and the HI-selected catalogs, $P_{\rm lin}(k)$ is the linear matter power spectrum convolved with the window corresponding to the shape of lightcone and $P_X^{\rm SN}$ is the shot noise given by~\citep{Wolz2017}
\begin{eqnarray}
    P_X^{\rm SN} = \frac{1}{\bar{n}_{\rm opt}}.
\end{eqnarray}
where $\bar{n}_{\rm opt}$ is the mean number density of the optical-selected catalog.
\begin{figure}[t]
    \centering
    \includegraphics[width=0.8\linewidth]{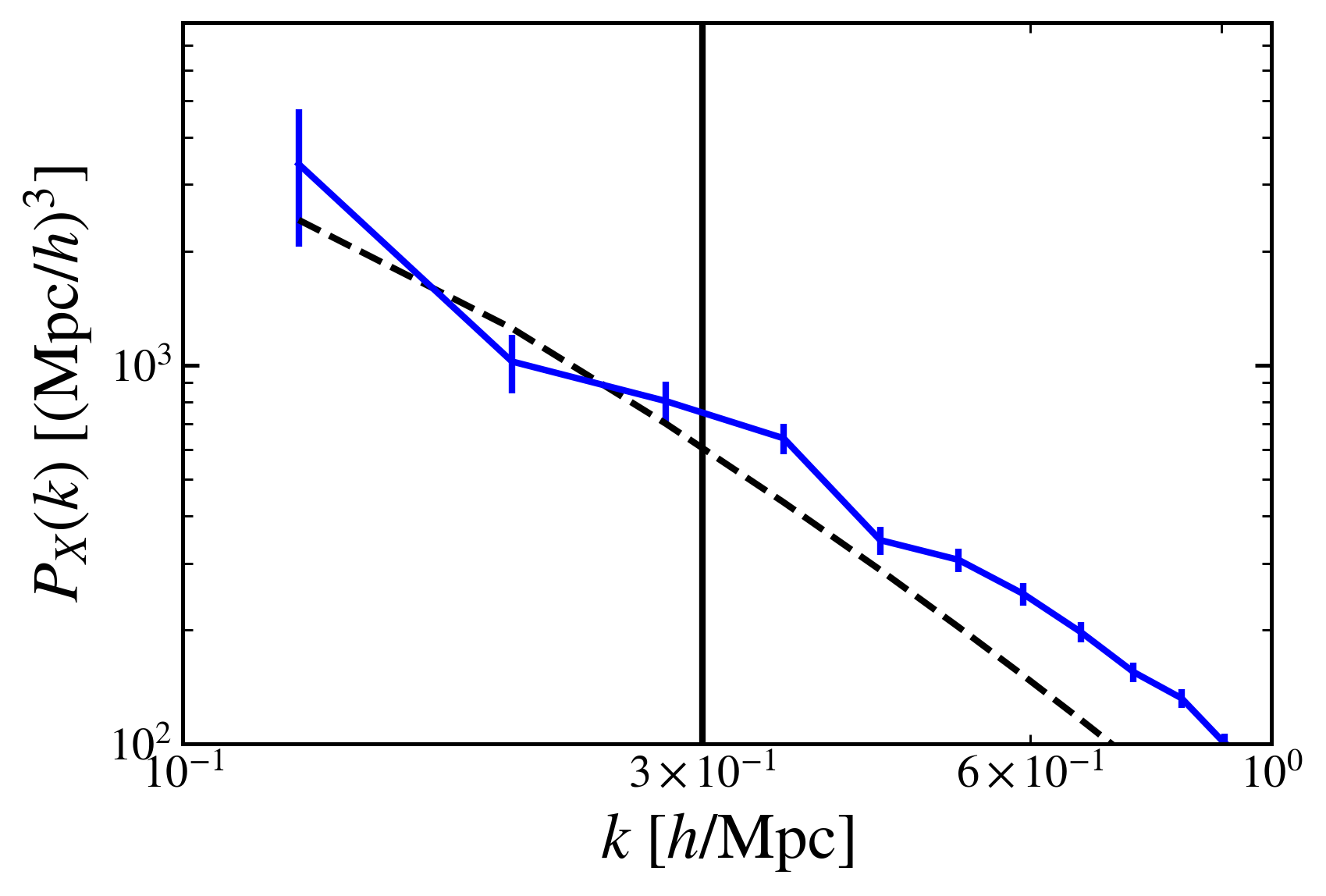}
    \caption{
        The cross power spectrum between the optical-selected catalog and the HI-selected catalog (blue solid line) at the redshift bin centered at 0.36.
        We also show the linear matter power spectrum at the same redshift.
        The vertical solid line indicate $k=0.3\,h/{\rm Mpc}$.
    }\label{fig:pkx}
\end{figure}

The error for cross power spectrum measurement is given by
\begin{eqnarray}
    &&\sigma_{P}(k) = \frac{1}{\sqrt{2N_{\rm mode}}}\sqrt{P_{\rm HI}(k)P_{\rm opt}(k) + P_X^2(k)},
\end{eqnarray}
where $N_{\rm mode}$ is the number of modes in a given $k$ bin, $P_{\rm HI}(k)$ and $P_{\rm opt}(k)$ are the total auto power spectrum (including shot noise) of the HI-selected catalog and optical-selected catalog, which we calculate directly from the lightcone.
The linear bias of the optical-selected catalog can be constrained to high precision, thanks to the large survey area of CSST,  so we neglect its measurement error and calculate it from the auto power spectrum of optical-selected catalog in the lightcone
\begin{equation}
    P_{\rm opt}(k) = b_{\rm opt}^2 P_{\rm lin}(k) + \frac{1}{\bar{n}_{\rm opt}}.
\end{equation}

The cross power spectrum at the redshift bin centered at $z=0.36$ is shown in Figure~\ref{fig:pkx}, along with the corresponding linear matter power spectrum.
\begin{figure}
    \centering
    \includegraphics[width=0.8\linewidth]{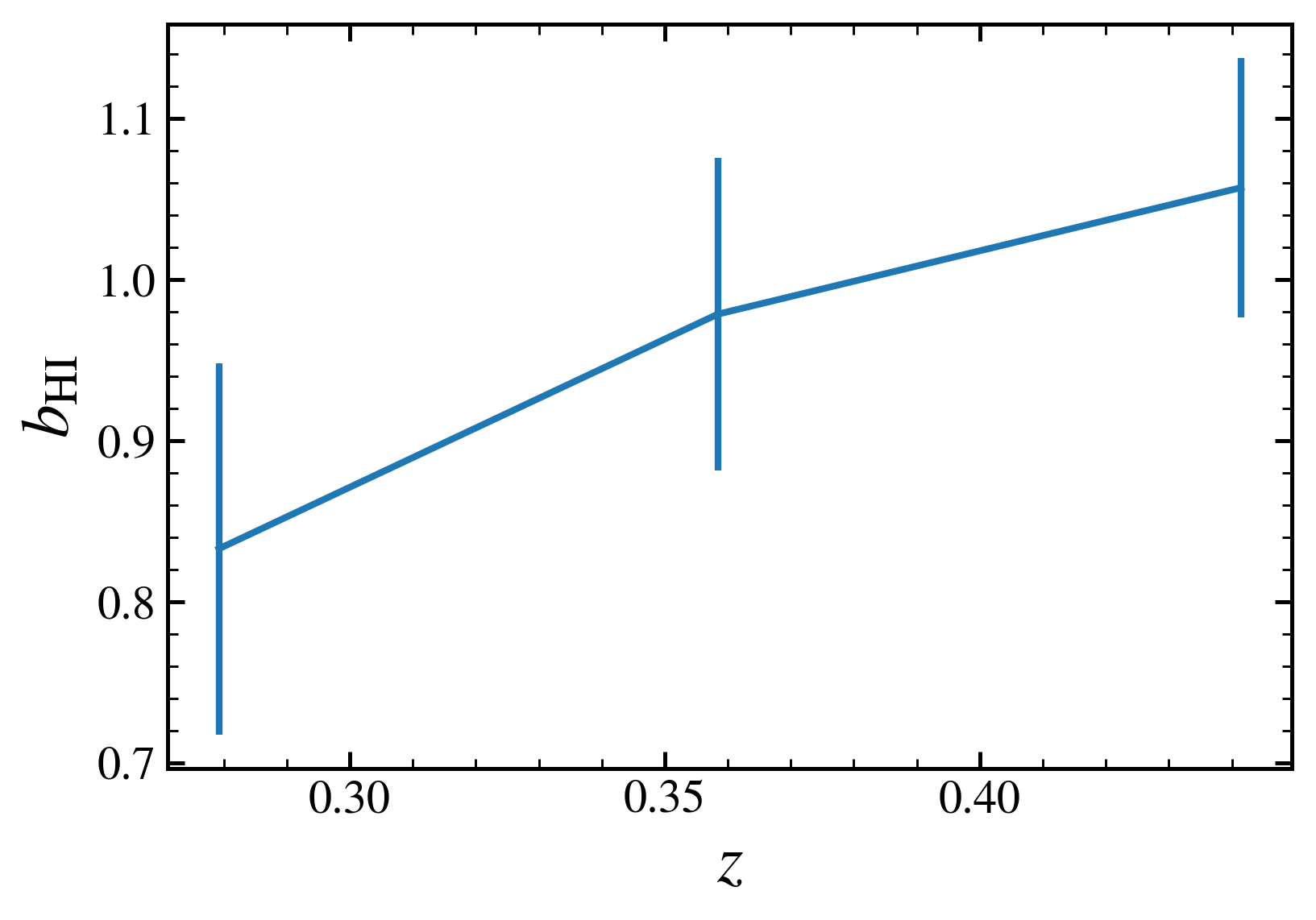}
    \caption{The linear bias of \HI\ $b_{\rm HI}$ at different redshift bins.}\label{fig:bHI}
\end{figure}

Using the cross power spectrum, we can estimate the bias of the \HI\ galaxies, and one can then use them to constrain the models of \HI\ galaxies (see e.g.~\citealt{Ibitoye2025}).
The cross power spectrum deviates from the linear matter power spectrum at $k\gtrsim 0.3\,h/{\rm Mpc}$ due to non-linear clustering, and therefore we use modes with $k< 0.3\,h/{\rm Mpc}$ to estimate $b_{\rm HI}$.
The \HI\ bias $b_{\rm HI}$ of the three redshift bins is shown in Figure~\ref{fig:bHI}.
The \HI\ galaxies have relatively small bias, showing that they trace the relatively low density regions.
The bias increases with the mean redshift.
This is because although we have used the more restrictive selection criteria to obtain samples with nearly constant \HI\ mass over the redshifts, the host halo mass still evolves with redshift, and at higher redshifts the corresponding bias is higher.

\section{Conclusion}\label{sec:conclusion}

In this paper we study the cross-correlation of galaxies detected in the optical band by the CSST spectroscopic survey, and the galaxies detected by the SKA-Mid in an \HI\ galaxy redshift survey.
We first construct the lightcone over the redshift range of $0.235<z<0.495$ over a 20 square degree field of view, using a semi-analytic model based on the  Millennium-II simulation, which has a box size of $95 \Mpc/h$ and a mass resolution of $7.7 \times 10^6 \Msun/h$. 
We test the detectability of the CSST and SKA.
For the SKA \HI\ survey, we consider one with a total integration time of 2000 hours, while for the CSST emission-line galaxy redshift survey, a total number of 4 observations is assumed, each with an exposure time of 150 s.

We generate the SKA \HI\ data cube according to the prescription of the SKA data challenge 2. 
The mock sample of the \HI\ galaxies is obtained by running the \SOFIA~\citep{Westmeier2021} source finding program.
The optical galaxies are selected to have ${\rm SNR}_{\rm eff}\geq 5$ for at least one of the strong emission-lines.
These represent realistic survey parameters.
We find that the SKA-Mid \HI\ survey SNR is basically proportional to the \HI\ mass of the galaxy, while the CSST emission-line SNR is proportional to the SFR.
The sample of galaxies detected in the \HI\ survey exhibits the BTFR, with the total baryonic (stellar and gas) mass proportional to the maximum rotation velocity.
By a stacking analysis of the galaxies detected in the CSST survey, the average \HI\ content can be effectively extracted.

Finally, we propose a selection criterion to identify subsamples of redshift-independent emission-line galaxies detectable by both CSST and SKA, and compute their cross power spectrum.
The measured cross-correlation power spectrum yields a measurement of the bias of the \HI\ galaxies, thus provide useful information about the \HI\ galaxies at low redshifts and help constrain the cosmology and galaxy evolution models.

\normalem
\begin{acknowledgements}
We thank the referee for comments and suggestions that improved the paper.
This work is supported by the National SKA Program of China (No. 2022SKA0110100, 2022SKA0110201), the National Natural Science Foundation of China (NSFC) International (Regional) Cooperation and Exchange Project (No. 12361141814), the NSFC Innovation group grant (No. 12421003), the National Key Research and Development Program of China (No. 2023YFA1607904), the National Natural Science Foundation of China (NSFC, grant No. 12503008), the China Manned Space Program with grant No. CMS-CSST-2025-A07.
The project is also supported by the Youth Innovation Promotion Association of the Chinese Academy of Sciences (No. 2022056), the Specialized Research Fund for State Key Laboratory of Radio Astronomy and Technology, the National Astronomical Observatories, Chinese Academy of Science (Nos. E5ZB0901), the CAS Project for Young Scientists in Basic Research (No. YSBR- 092), National Key R\&D Program of China grant Nos. 2022YFF0503404, and the science research grant from the China Manned Space Project with grant Nos. CMS- CSST-2025-A02.
\end{acknowledgements}

\bibliographystyle{raa}
\bibliography{cross}

\end{document}